\documentclass[11pt]{article}

\usepackage{graphicx}
\usepackage{amsmath}
\usepackage{amssymb}
\usepackage{hyperref}
\usepackage{cite}
\usepackage{times}
\usepackage{fullpage}
\usepackage{ifthen}
\usepackage{deluxetable}

\newcommand{\spacing}[1]{\renewcommand{\baselinestretch}{#1}\large\normalsize}
\spacing{1}

\def\@maketitle{%
  \newpage\spacing{1}\setlength{\parskip}{12pt}%
    {\Large\bfseries\noindent\sloppy \textsf{\@title} \par}%
    {\noindent\sloppy \@author}%
}

\def\arcsec{\hbox{$^{\prime\prime}$}}
\def\arcmin{\hbox{$^{\prime}$}}
\def\degr{\hbox{$^{\circ}$}}
\newcommand{\farcs}{\mbox{\ensuremath{.\!\!^{\prime\prime}}}}

\newcommand{\physrep}{{Phys. Rep.}}
\newcommand{\aapr}{{Astron. Astrophys. Rev.}}
\newcommand{\mnras}{{Mon. Not. R. Astron. Soc.}}
\newcommand{\apj}{{Astrophys. J.}}

\newcommand{\aap}{{Astron. Astrophys.}}
\newcommand{\apjs}{{Astrophys. J. Suppl. Ser.}}
\newcommand{\aj}{{Astron. J.}}
\newcommand{\apjl}{{Astrophys. J.}}
\newcommand{\pasj}{{Publ. Astron. Soc. Jpn.}}
\newcommand{\procspie}{{Proc. SPIE Int. Soc. Opt. Eng.}}
\newcommand{\apss}{{Astrophys. Space Sci.}}

\newcommand{\RNum}[1]{\uppercase\expandafter{\romannumeral #1\relax}}

\newcommand\ion[2]{#1$\;${\RNum{#2}}}%

\newenvironment{addendum}{%
    \setlength{\parindent}{0in}%
    \small%
    \begin{list}{Acknowledgements}{%
        \setlength{\leftmargin}{0in}%
        \setlength{\listparindent}{0in}%
        \setlength{\labelsep}{0em}%
        \setlength{\labelwidth}{0in}%
        \setlength{\itemsep}{12pt}%
        }
    }
    {\end{list}\normalsize}

\newenvironment{methods}{%
    \section*{Methods}%
    \setlength{\parskip}{6pt}%
    }{}

\title{The Case for Electron Re-Acceleration at Galaxy Cluster Shocks\vspace{-2cm}}

\date{}

\begin{document}

\maketitle

\newenvironment{affiliations}{%
    \setcounter{enumi}{1}%
    \setlength{\parindent}{0in}%
    \slshape\sloppy%
    \begin{list}{\upshape$^{\arabic{enumi}}$}{%
        \usecounter{enumi}%
        \setlength{\leftmargin}{0in}%
        \setlength{\topsep}{0in}%
        \setlength{\labelsep}{0in}%
        \setlength{\labelwidth}{0in}%
        \setlength{\listparindent}{0in}%
        \setlength{\itemsep}{0ex}%
        \setlength{\parsep}{0in}%
        }
    }{\end{list}\par\vspace{12pt}}

\renewenvironment{abstract}{%
    \setlength{\parindent}{0in}%
    \setlength{\parskip}{0in}%
    \bfseries%
    }{\par\vspace{0pt}}


{\bf
{\noindent \author{Reinout J. van Weeren$^{1}$, Felipe~Andrade-Santos$^{1}$, William~A.~Dawson$^{2}$,
 Nathan~Golovich$^{3}$, Dharam~V.~Lal$^{4}$, Hyesung Kang$^{5}$, Dongsu Ryu$^{6,7}$, 
Marcus Br\"uggen$^{8}$, Georgiana~A.~Ogrean$^{9}$, William~R.~Forman$^{1}$, Christine~Jones$^{1}$, 
Vinicius~M.~Placco$^{10}$, Rafael~M.~Santucci$^{11}$, David~Wittman$^{3,12}$, 
M.~James~Jee$^{13}$, Ralph~P.~Kraft$^{1}$, David~Sobral$^{14,15}$, 
Andra~Stroe$^{16}$ \& Kevin~Fogarty$^{17}$}
}
}


\begin{affiliations}
\item Harvard-Smithsonian Center for Astrophysics, 60 Garden Street, Cambridge, MA 02138, USA
\item Lawrence Livermore National Lab, 7000 East Avenue, Livermore, CA 94550, USA
\item University of California, One Shields Avenue, Davis, CA 95616, USA
\item National Centre for Radio Astrophysics, TIFR, Pune University Campus, Post Bag 3, Pune 411007, India
\item Department of Earth Sciences, Pusan National University, Busan 46241, Korea
\item Department of Physics, UNIST, Ulsan 44919, Korea
\item Korea Astronomy and Space Science Institute, Daejeon 34055, Korea
\item Hamburger Sternwarte, Hamburg University, Gojenbergsweg 112, 21029 Hamburg, Germany
\item Kavli Institute for Particle Astrophysics and Cosmology, Stanford University, 452 Lomita Mall, Stanford, CA 94305-4085, USA
\item Department of Physics and JINA Center for the Evolution of the Elements, University of Notre Dame, Notre Dame, IN 46556, USA
\item Departamento de Astronomia - Instituto de Astronomia, Geof\'isica e Ci\^encias Atmosf\'ericas, Universidade de S\~ao Paulo, S\~ao Paulo, SP 05508-900, Brazil
\item Instituto de Astrof\'isica e Ci\^encias do Espa\c{c}o, Universidade de Lisboa, Lisbon 1749-016, Portugal
\item Department of Astronomy and Center for Galaxy Evolution Research, Yonsei University, 50 Yonsei-ro, Seoul 03722, Korea
\item Department of Physics, Lancaster University, Lancaster LA1 4YB, UK
\item Leiden Observatory, Leiden University, PO Box 9513, NL-2300 RA Leiden, the Netherlands
\item European Southern Observatory, Karl-Schwarzschild-Stra{\ss}e 2, D-85748 Garching bei M\"uchen, Germany
\item Department of Physics and Astronomy, The Johns Hopkins University, 3400 North  Charles Street, Baltimore, MD 21218-2686, USA
\end{affiliations}

\begin{abstract}

On the largest scales, the Universe consists of voids and filaments making up the cosmic web. Galaxy clusters are located at the knots in this web, at the intersection of filaments. Clusters grow through accretion from these large-scale filaments and by mergers with other clusters and groups.
In a growing number of galaxy clusters, elongated Mpc-size radio sources have been found, so-called radio relics \cite{2012A&ARv..20...54F,2014IJMPD..2330007B}. These relics are thought to trace relativistic electrons in the intracluster plasma accelerated by low-Mach number collisionless shocks generated by cluster-cluster merger events \cite{1998A&A...332..395E}. A long-standing problem is how low-Mach number shocks can accelerate electrons so efficiently to explain the observed radio relics. Here we report on the discovery of a direct connection between a radio relic and a radio galaxy in the merging galaxy cluster Abell~3411-3412. This discovery indicates that fossil relativistic electrons from active galactic nuclei are re-accelerated at cluster shocks. It also implies that radio galaxies play an important role in governing the non-thermal component of the intracluster medium in merging clusters.
\end{abstract}

Cluster mergers are the most energetic events in the Universe after the Big Bang, releasing energies up to $\sim10^{64}$ ergs on Gyr timescales.
Most of the gravitational energy released during cluster merger events is converted into thermal energy via low-Mach number shocks  ($\mathcal{M} \lesssim 3$) and turbulence in the intracluster medium (ICM) \cite{2003ApJ...593..599R}. A small fraction ($\lesssim 1$\%) of the energy dissipated at shocks could be channeled into the acceleration of cosmic rays (CR). In the presence of magnetic fields,  CR electrons would then emit synchrotron radiation which can be observed with radio telescopes. The origin of the large-scale magnetic fields, and the nature of particle acceleration processes that operate in these dilute cosmic plasmas, are still open questions.

The ICM has a high thermal-to-magnetic pressure ratio, or $\beta$, and electron acceleration by low-Mach number collisionless shocks in such high-$\beta$ plasmas is poorly understood, as analytical calculations cannot properly capture the non-linear behavior of this process \cite{2011Ap&SS.336..263K}. Radio relics, elongated sources that trace the CR at ICM shocks, provide us with rare opportunities to probe this process. While there is substantial evidence that relics trace CR electrons at shocks \cite{2011ApJ...728...82M,2015MNRAS.449.1486S}, previous work has found that the acceleration efficiency should be very low at these shocks, if these synchrotron emitting electrons are accelerated from the thermal pool of the ICM via the diffusive shock acceleration (DSA) mechanism \cite{2011Ap&SS.336..263K}. This low efficiency is hard to reconcile with the observed brightness of some radio relics, suggesting a high acceleration efficiency \cite{2012ApJ...756...97K,2013MNRAS.435.1061P,2014MNRAS.437.2291V}. In addition, some relics have regions with rather flat radio spectra ($\alpha\approx -0.7$; $F_\nu \propto \nu^{\alpha}$, where $\alpha$ is the spectral index), but the corresponding shock Mach numbers measured via X-ray observations are low \cite{2015PASJ...67..113I,2016ApJ...818..204V}. This contradicts with the prediction from DSA \cite{1987PhR...154....1B}.  This long-standing problem has so far remained unsolved. Furthermore, large merger shocks have also been found without corresponding radio relics\cite{2011MNRAS.417L...1R},  indicating that our understanding of particle acceleration by low-Mach number shocks is still incomplete. 

Recently,  new insights into the acceleration by low-Mach number shocks have been obtained by particle-in-cell (PIC) simulations \cite{2014ApJ...797...47G}. These PIC simulations show that  acceleration from the thermal pool could be possible. However, for some relics an unrealistic fraction of the shock energy needs to be transferred into the non-thermal electron population to explain their radio brightness \cite{2016ApJ...818..204V,2016MNRAS.460L..84B,2016MNRAS.461.1302E} and the PIC simulations do not solve this problem.
A solution to explain the apparent very efficient
acceleration, is to invoke the presence of a population of
fossil relativistic electrons \cite{2005ApJ...627..733M}, with Lorentz factors $\gamma \gtrsim 10^2$. The synchrotron lifetime of relativistic 
electrons in the ICM is relatively short ($\sim 10^{8}$~yrs). Once
these electrons have lost most of their energy, they do not radiate
within the observable radio band and they thus become invisible to radio telescopes. It has been suggested that these fossil electrons, which have Gyr lifetimes,  can be efficiently re-accelerated at shocks and are therefore
able to create bright radio relics \cite{2013MNRAS.435.1061P,2015ApJ...809..186K}. Obvious candidates
for these fossil electrons are the (old) lobes and tails of radio
galaxies \cite{2014ApJ...785....1B,2015MNRAS.449.1486S,1991A&A...252..528G}. This fossil radio plasma can occupy a significant volume of the ICM due to turbulent diffusion, aided by
the random galaxy motions throughout the ICM and the ICM motion itself. Indeed, observational evidence has been found that provides support for this model. The complex morphologies of some radio relics resemble those of disturbed tailed radio galaxies that are often found in merging galaxy clusters\cite{2014ApJ...785....1B,2015MNRAS.449.1486S}. The relics in the Bullet Cluster and PLCKG287.0+32.9 are two prime examples \cite{2015MNRAS.449.1486S, 2011ApJ...736L...8B}.
However, for the cluster PLCKG287.0+32.9, no redshift for the putative radio galaxy connected to the relic could be obtained, since the proposed core of the radio galaxy did not have an optical counterpart \cite{2014ApJ...785....1B}. Similarly,  no direct link between the relic in the Bullet cluster and a tailed radio galaxy could be established \cite{2015MNRAS.449.1486S}. The relic in the Coma cluster provides another interesting case. Here a confirmed cluster radio galaxy seems to be connected to a relic \cite{1991A&A...252..528G}. Adiabatic compression of fossil plasma by a shock was proposed as a model for its origin \cite{2002MNRAS.331.1011E,2001A&A...366...26E}. While the electrons do gain energy by the adiabatic compression, this model does not invoke any Fermi-type re-acceleration processes.

\begin{figure*}[h!]
    \begin{center}
          \includegraphics[angle = 0, trim =0cm 0cm 0cm 0cm,width=0.75\textwidth]{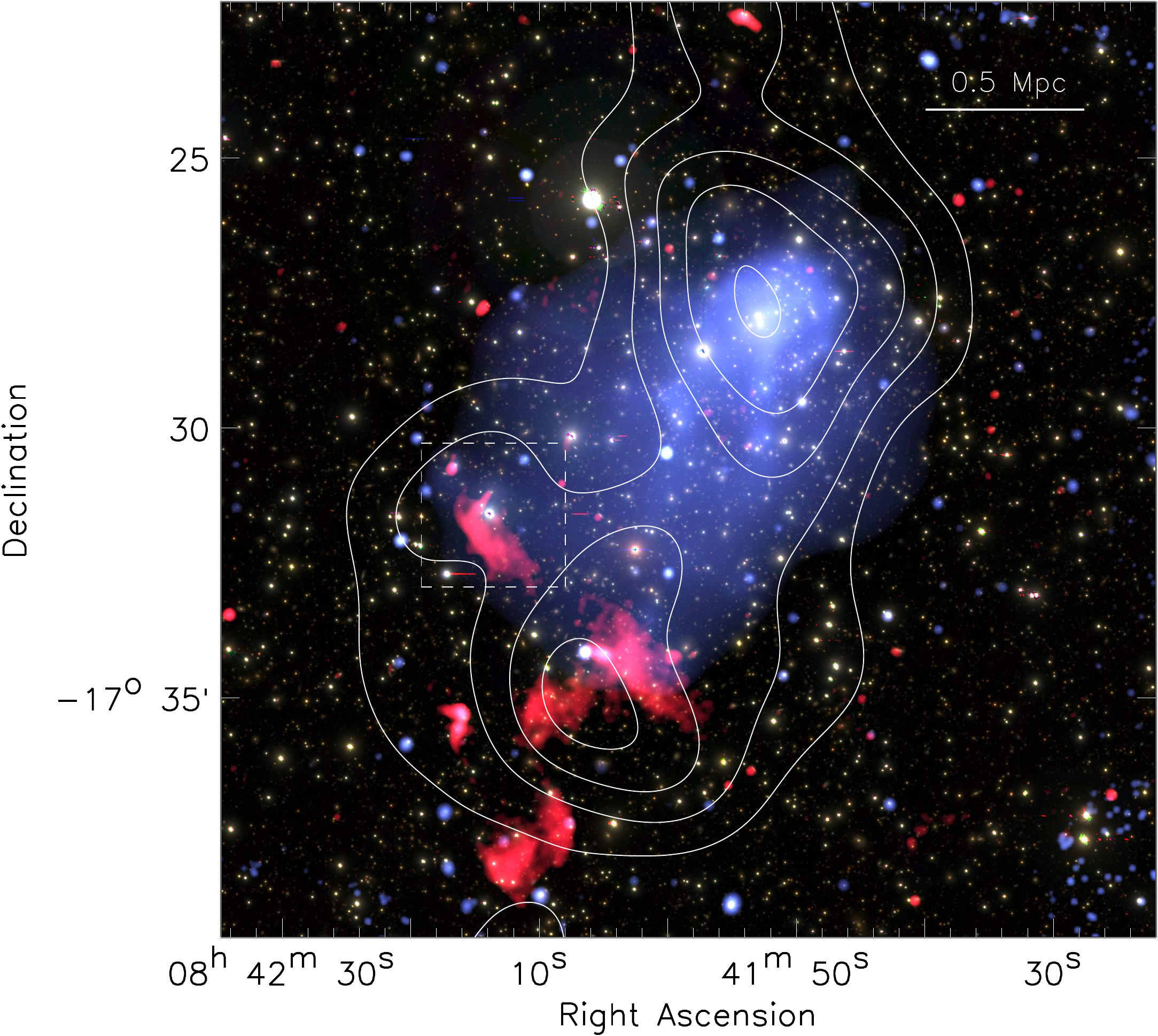}
       \end{center}
      \caption{Subaru \emph{gri} color image of the merging cluster Abell~3411-3412. Radio emission at 610~MHz from the GMRT is shown in red. The 0.5--2.0~keV Chandra X-ray image is shown in blue. The galaxy distribution is shown with white contours. Contour levels are drawn at $[3,4,5,\ldots]$~galaxies~arcmin$^{-2}$. The dashed box indicates the region shown in Fig.~\ref{fig:spix}.}
      \label{fig:mergeroverview}
 \end{figure*}

Here we present optical, radio, and X-ray observations of the merging galaxy cluster Abell~3411-3412 located at $z=0.162$. This cluster contains a Mpc-size radio relic with an irregular shape \cite{2013ApJ...769..101V, 2013MNRAS.435..518G}. Chandra observations, totaling 211~ks, reveal a cluster with a cometary morphology undergoing a major merger event, with the compact core of one of the subclusters being the ``head'' of the comet (Fig.~\ref{fig:mergeroverview}). A dynamical analysis based on Keck spectra of 174  cluster members and Subaru imaging, indicates that this is an approximately 1:1 mass ratio merger viewed $\sim 1$~Gyr after core passage, with the merger axis being located close to the plane of the sky. The two subclusters are both very massive, with individual masses of $\sim 10^{15}$~M$_{\odot}$. The core of the cluster coming in from the south (and currently observed in the north) was compact enough to survive  the collision with the other cluster up to the present time, while the gas core of the subcluster that came in from the north was largely disrupted during the core passage.

\begin{figure*}[h!]
    \begin{center}
           \includegraphics[angle = 180, trim =0cm 0cm 0cm 0cm,width=0.49\textwidth]{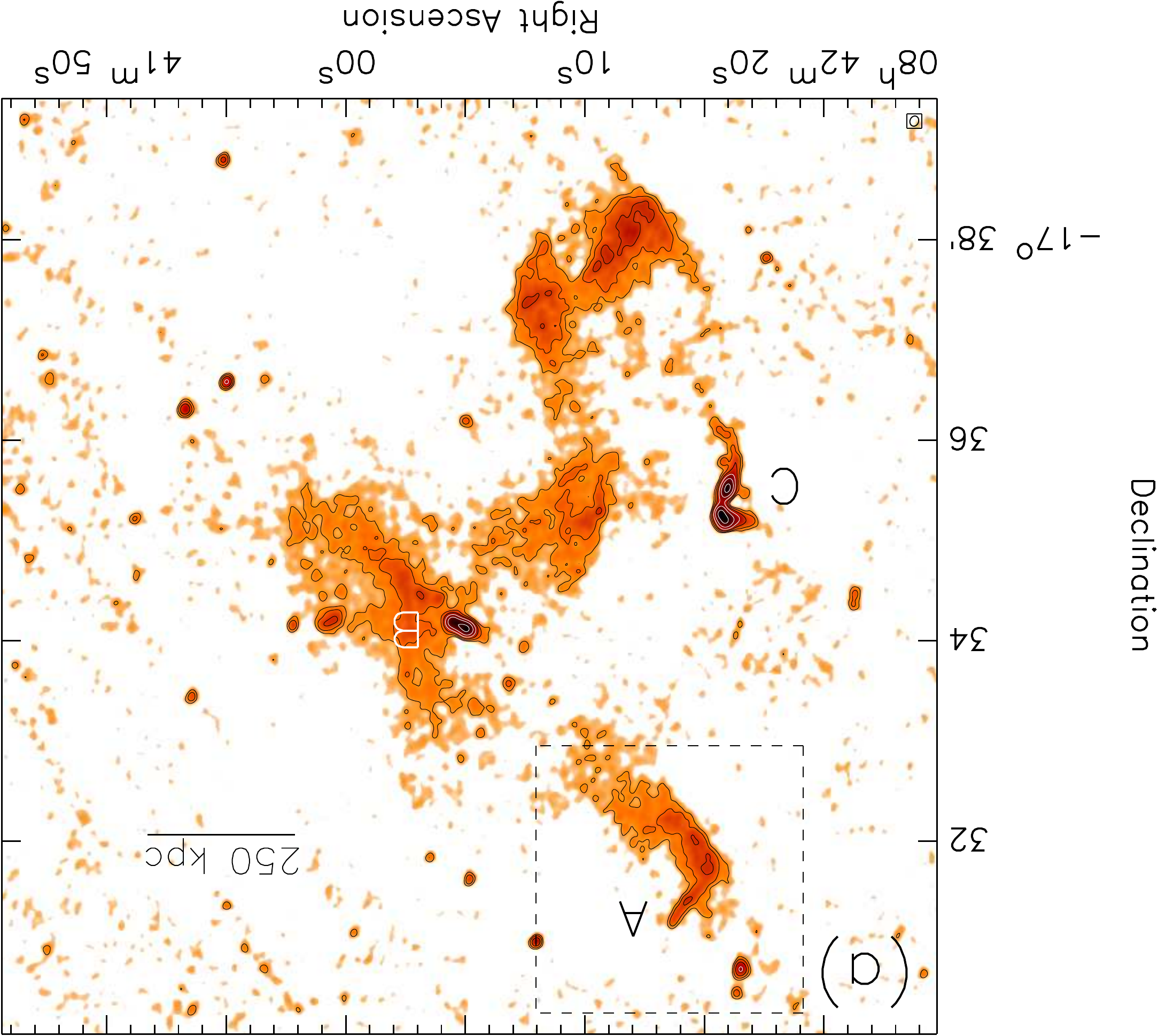}
                \includegraphics[angle = 180, trim =0cm 0cm 0cm 0cm,width=0.49\textwidth]{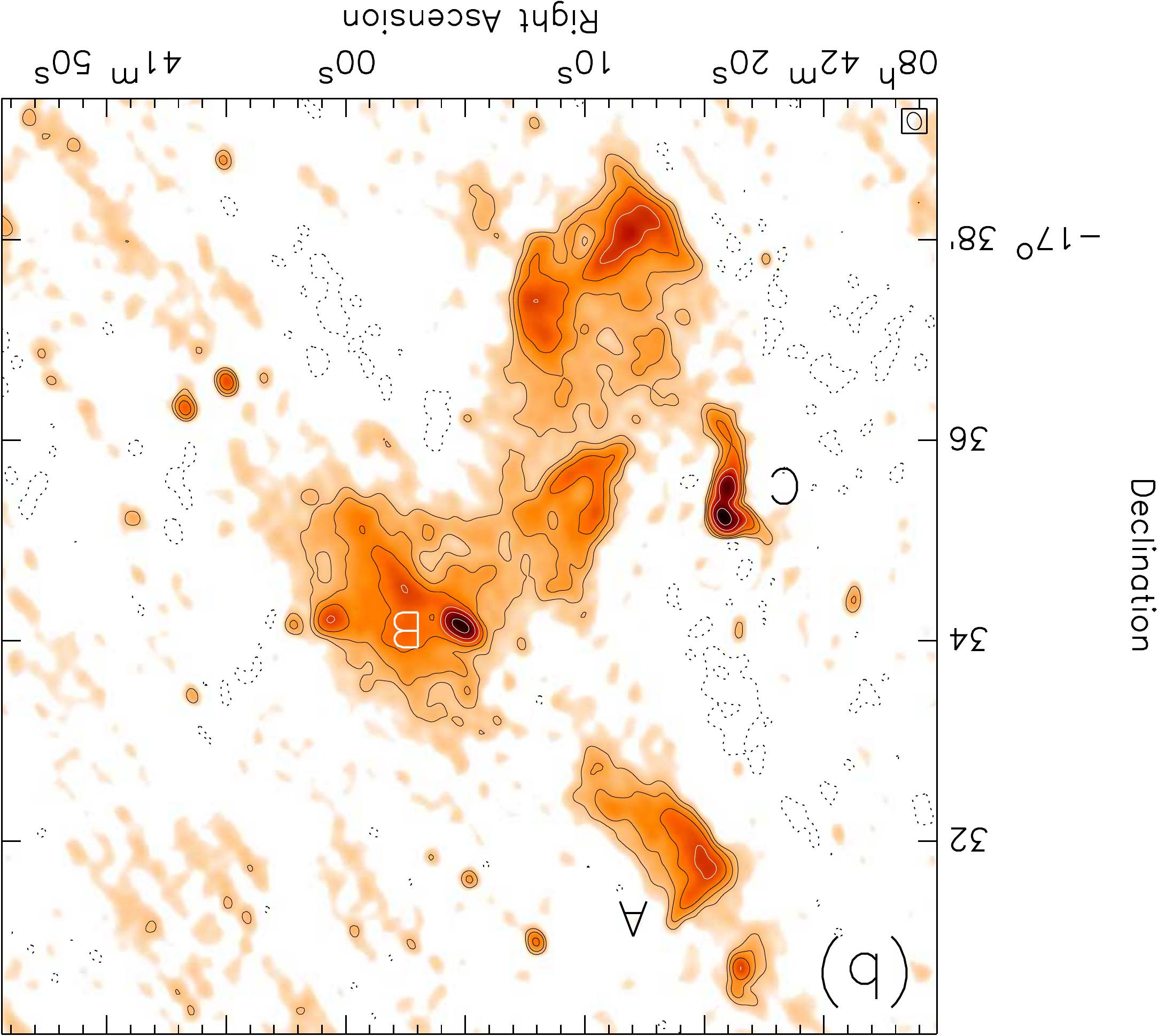}
       \end{center}
       \caption{GMRT radio images. (\emph{a}) GMRT 610~MHz image. Contours are drawn at levels of $\sqrt{[1,2,4, \ldots]} \times 4\sigma_{\rm{rms}}$, with $\sigma_{\rm{rms}}$ the map noise. The beam size is indicated in the bottom left corner. The dashed box indicates the region shown in Fig.~\ref{fig:spix}. Cluster radio galaxies are labeled A, B, and C. (\emph{b}) GMRT~325~MHz image. Contour levels are drawn as in the left panel. Dotted contours are drawn at $-3\sigma_{\rm{rms}}$.}
      \label{fig:gmrt610}
 \end{figure*}

\begin{figure*}[h!]
    \begin{center}
          \includegraphics[angle = 180, trim =0cm 0cm 0cm 0cm,width=0.3\textwidth]{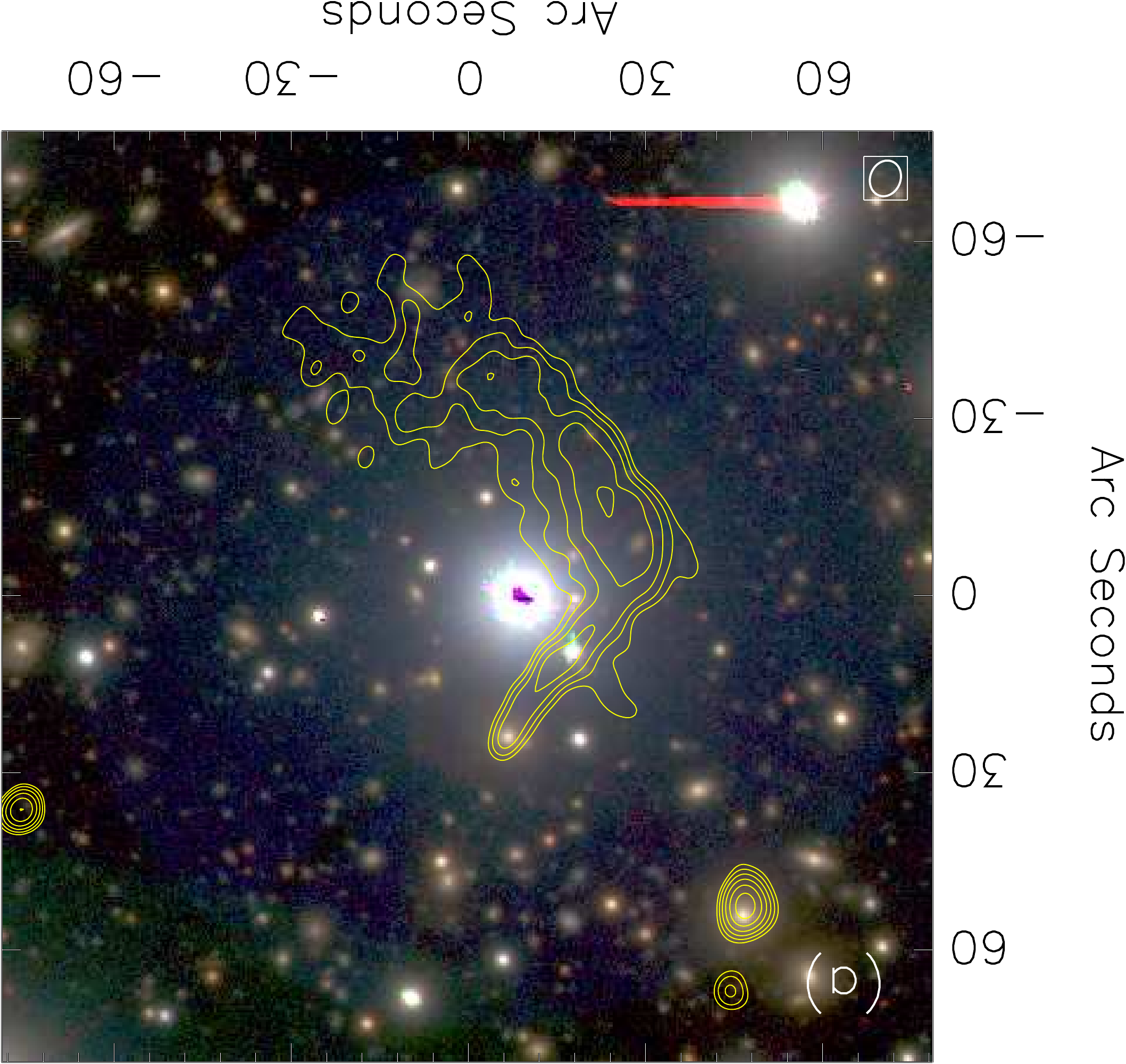}
            \includegraphics[angle = 180, trim =0cm 0cm 0cm 0cm,width=0.3\textwidth]{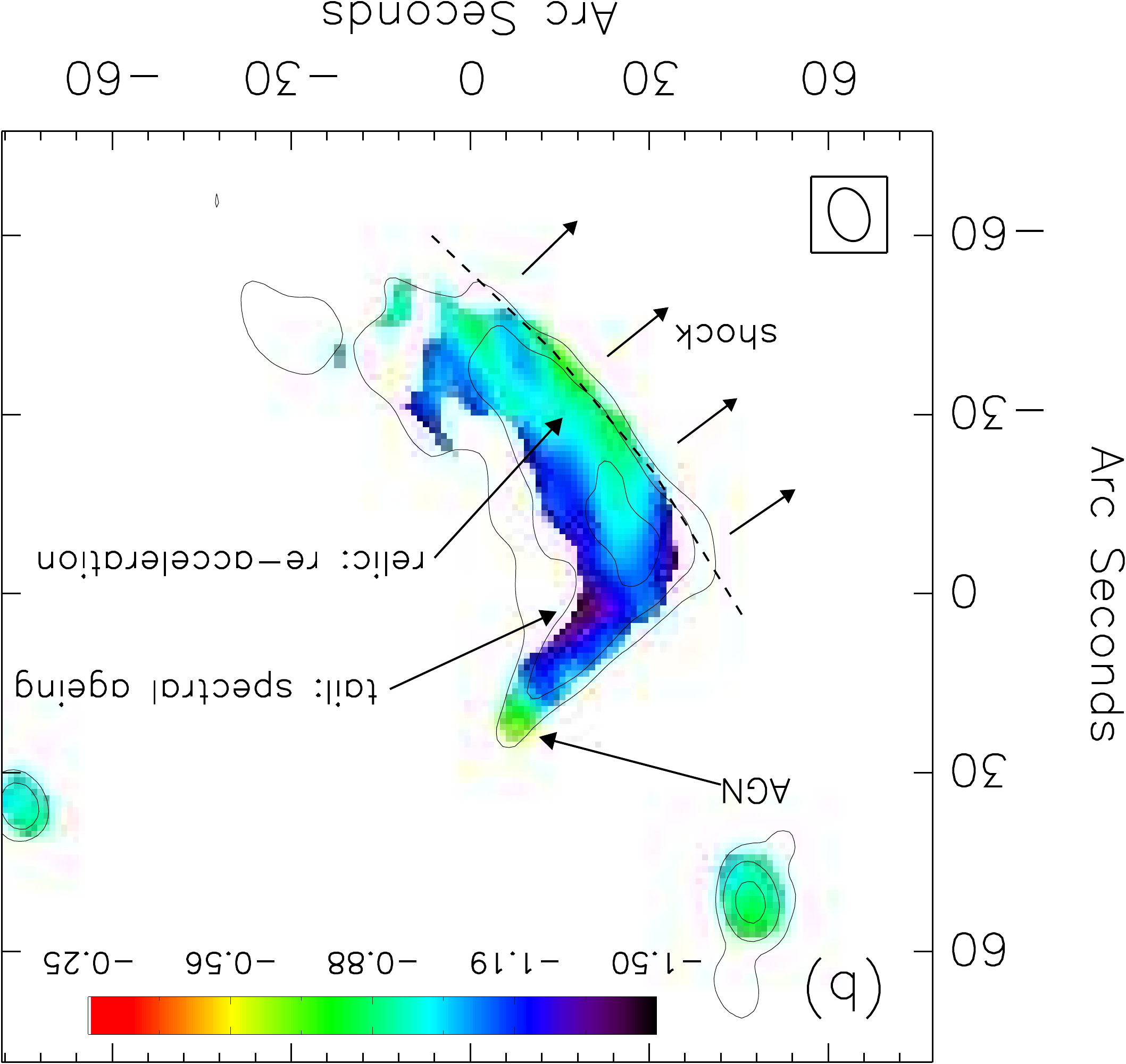}
         \includegraphics[angle = 180, trim =0cm 0cm 0cm 0cm,width=0.3\textwidth]{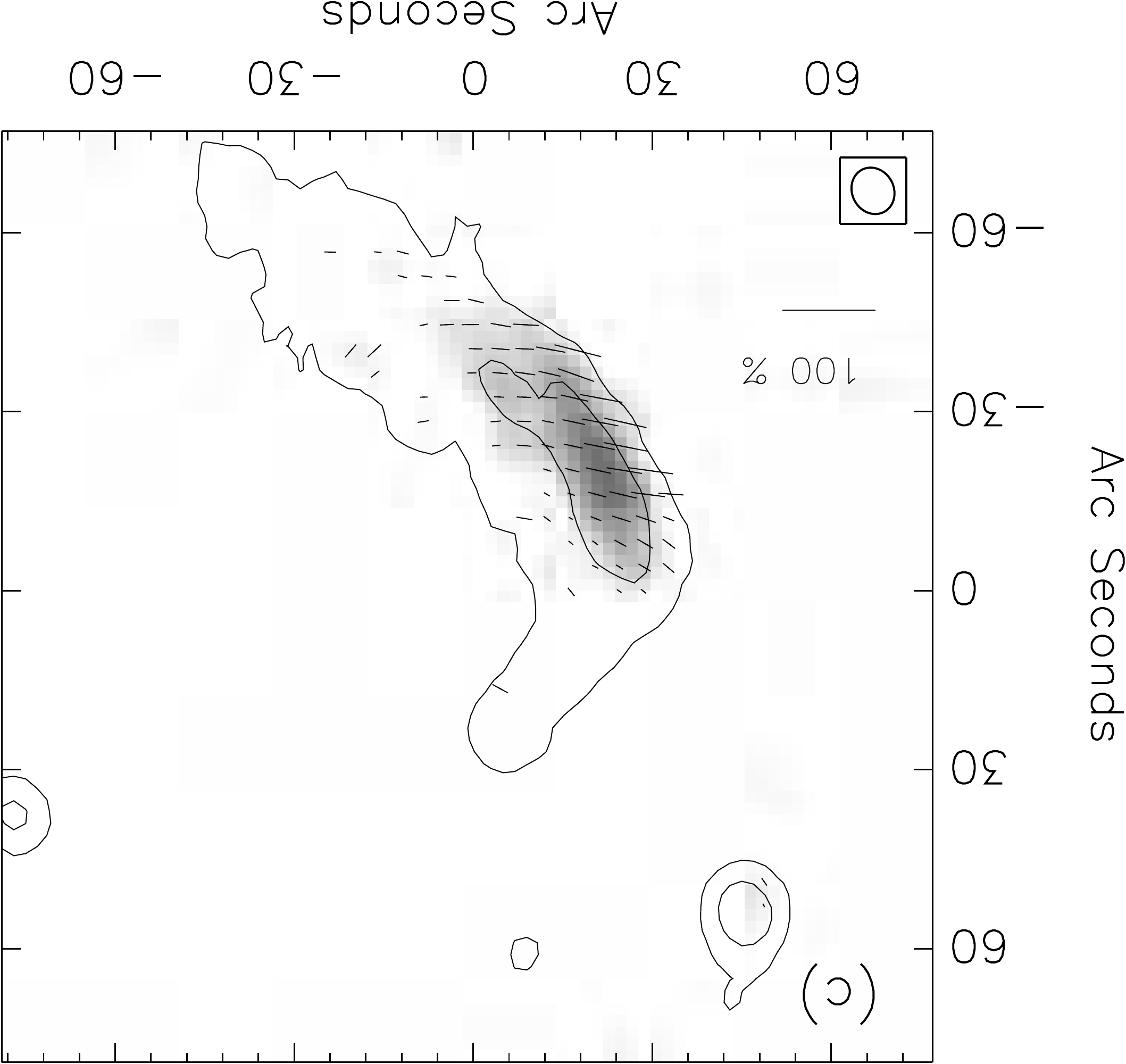}
     \end{center}
              \caption{Subaru optical, radio spectral index, and polarization maps of the of the northeast component of the radio relic in Abell~3411-3412. The beam sizes are indicated in the bottom left corner of the images. 
       (\emph{a}) Subaru  \emph{gri} color image. GMRT 610~MHz contours are overlaid in yellow. Contours are drawn at levels of $\sqrt{[1,2,4,\ldots]} \times 5\sigma_{\rm{rms}}$. 
       (\emph{b}) A spectral index map. The spectral index was determined using matched observations at frequencies of 0.325, 0.61, 1.5 and 3.0~GHz. Contours are from the GMRT 325~MHz image and are drawn at levels of ${[1,2,4,\ldots]} \times 6\sigma_{\rm{rms}}$. The dashed line indicates the best-fitting position of the X-ray surface brightness edge.
       (\emph{c}) Polarization map at 3.0~GHz. Vectors display the electric field angles. The length of the vectors is proportional to the polarization fraction. A reference vector for 100\% polarization is shown in the bottom left corner. The greyscale image shows the linear polarized intensity. Black contours trace the Stokes~I continuum image and are drawn at $[0.03,0.12]$~mJy~beam$^{-1}$. }
      \label{fig:spix}
 \end{figure*}

The Giant Metrewave Radio Telescope (GMRT) radio images at 610 and 325~MHz show the complex, large radio relic located in the southern outskirts of the merging cluster (Fig.~\ref{fig:gmrt610}). Most interestingly, a close inspection of the northeastern component of the radio relic reveals an elongated narrow extension that leads from the relic towards a galaxy (source~A, Figs.~\ref{fig:gmrt610} and \ref{fig:spix}a). This galaxy is a spectroscopically confirmed cluster member ($z=0.164$). A high-resolution 3~GHz Very Large Array (VLA) image shows an active galactic nucleus (AGN; radio core) at the galaxy's center. From the core, a narrow tail ``feeds'' into the radio relic located $\approx90$~kpc to its south. This reveals a direct connection between the relativistic plasma of an AGN and that of a radio relic. The probability of a chance projection of a tailed radio galaxy with this relic is $0.004$.

Our radio spectral index measurements, between 0.325 and 3.0~GHz show that the radio spectrum steepens away with distance from the compact nucleus from $\alpha = -0.5\pm0.1$ to $ \alpha = -1.3\pm0.1$ ($1\sigma$ errors), as is expected for synchrotron and Inverse Compton losses (Fig.~\ref{fig:spix}b). Remarkably, where the radio plasma from the AGN connects to the radio relic, the spectrum flattens back to $\alpha = -0.9 \pm 0.1$. 
This flattening is strong evidence for the  re-acceleration of electrons from the radio tail at a shock. Moreover, we find evidence for spectral steepening across the relic in the direction towards the cluster center. These gradients have routinely been found for other relics and they are explained by electron energy losses in the downstream region of an outwards traveling shock \cite{2010Sci...330..347V}. We also find the outer edge of the relic is polarized, with a maximum  polarization fraction of $40\%$. The emission weighted average polarization fraction is 13\% (Fig.~\ref{fig:spix}c).  The relatively modest spectral flattening at the shock suggests that the shock cannot be very strong. Furthermore, the existence of a downstream spectral index gradient (see Fig.~\ref{fig:spix}b) suggests that the relic cannot be solely caused by the adiabatic compression of a  lobe of fossil radio plasma.

\begin{figure*}[h!]
    \begin{center}
          \includegraphics[angle = 0, trim =0cm 0cm 0cm 0cm,width=0.5\textwidth]{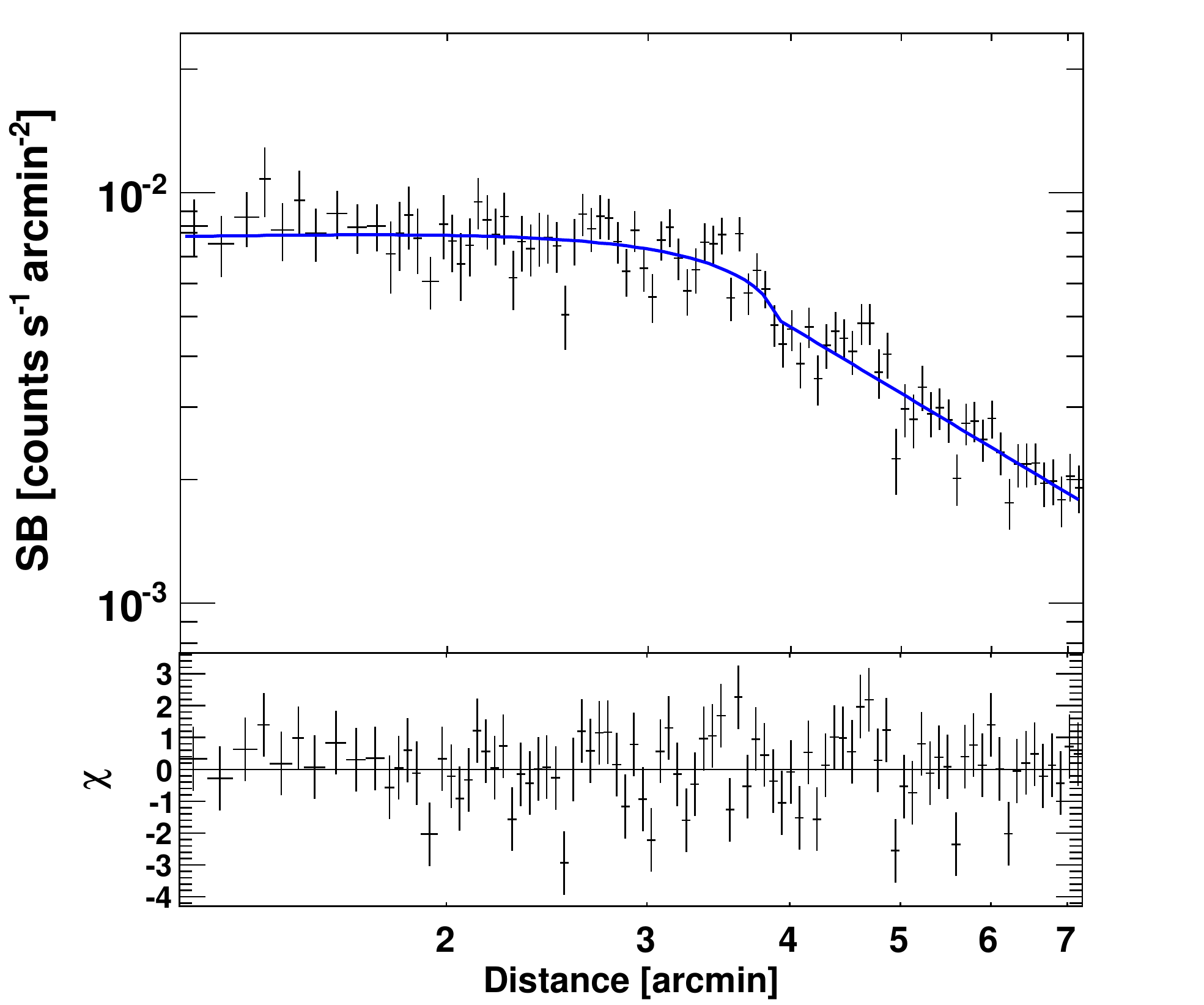}
        \end{center}
       \caption{Chandra 0.5--2.0~keV surface brightness profile across the radio relic in an elliptical sector (see the SI).  Uncertainties on the data points are plotted at the $1\sigma$ level. The blue line corresponds to the best-fitting density model. This double power-law model fits a jump (at a distance of $\approx 4\arcmin$) with a density compression factor ($C$) of $1.31^{+0.10}_{-0.09}$. The position of this jump is indicated on the spectral index image (Fig.~\ref{fig:spix}).}
       \label{fig:profile}
 \end{figure*}

To search for the shock, we extracted a 0.5--2~keV Chandra surface brightness profile in a sector containing the relic and fitted the profile by assuming an underlying broken power-law density model. At large radii, the density profiles of galaxy clusters are typically  described by single power-laws. However, the profile across the NE part of the relic steepens swiftly near the relic's outer edge and shows a significant  departure from a single power-law, indicating a deviation from hydrostatic equilibrium as expected for a shock (Fig.~\ref{fig:profile}). 
From the profile we find that the shock must be rather weak, with $\mathcal{M} \lesssim 1.7$. This upper limit should be considered an approximate estimate due to the idealized assumptions about the shock geometry and projection effects. 
 
Our observations of  (1) a direct connection between a radio galaxy and the relic, (2) spectral flattening at the location where the radio tail meets the relic, (3) the presence of an X-ray surface brightness discontinuity at the relic's outer edge, and (4) a high polarization fraction at the location of flattest spectral indices 
provide the best evidence to date that radio galaxies in clusters provide seed electrons that can be re-accelerated and revived by merger shocks. Re-acceleration (and also adiabatic compression) alleviates the problem of the low-acceleration efficiency from the thermal pool. In addition, it provides a natural explanation why some cluster merger shocks have no corresponding radio relics. Those cases may lack a sufficient supply of fossil radio plasma. Finally, re-acceleration also solves the problem of the relatively flat spectral indices observed from relics, which for DSA, requires shocks that have higher Mach numbers than observed in X-rays \cite{2015PASJ...67..113I,2016ApJ...818..204V}.  

Different re-acceleration models have been proposed.
{In one of these models, fossil relativistic electrons are re-accelerated by a DSA-like process \cite{2005ApJ...627..733M, 2015ApJ...815..116F} in combination with adiabatic compression. In another model, the re-acceleration happens in the shock downstream region by magneto-hydrodynamical turbulence \cite{2015ApJ...815..116F}.} Future work is needed to determine which of  these re-acceleration mechanisms operates at radio relics \cite{2015MNRAS.451.2198V}, and determine whether re-acceleration is required to explain all  relics. 
We evaluate one of these re-acceleration models in the \emph{Supplementary Information} to investigate what Mach number is required to flatten the spectral index by the observed amount and examine the relatively uniform spectral index along the length of the shock front.

The Abell~3411-3412 cluster contains at least two additional radio galaxies about 2\arcmin~to the south and 2\arcmin~to southwest of the one described above (sources~B and~C; Fig.~\ref{fig:gmrt610}). One of the radio galaxies is embedded within the relic emission, for the other, a tail of steep spectrum radio emission extends towards the relic. We argue that it is therefore likely that the other components of the complex relic in Abell~3411-3412 also trace revived fossil plasma, either by the process of adiabatic compression or by re-acceleration. Re-acceleration should also operate in other clusters because more examples of radio galaxies near relics have been found (i.e., in the Coma Cluster \cite{1991A&A...252..528G}). Our findings imply that PIC simulations and laboratory laser experiments\cite{laser} for collisionless shocks should include a relativistic fossil electron distribution in the upstream plasma. This study also indicates that to understand the non-thermal component of the ICM, the presence and distribution of radio galaxies needs to be taken into account, in addition to particle acceleration at shocks. Observations at low radio frequencies, in particular with LOFAR, will be key to unveiling the connections between relics and AGN, because low-frequency observations are sensitive to older, fossil radio plasma.


\clearpage

\begin{methods}

\subsubsection*{Radio observations, data reduction, and spectral indices}

Abell~3411-3412 was observed with the GMRT on 21~Oct and 19~Nov, 2012, at 610 and 325~MHz, respectively. A total bandwidth of 32~MHz was recorded. The on source time was 4.5~hrs at 325~MHz and 3.6~hrs at 610~MHz. The initial calibration and visual removal of radio frequency interference (RFI) were carried out with AIPS (\url{http://www.aips.nrao.edu/}). The initial calibration consisted of bandpass calibration, bootstrapping of the flux-density scale, and the transfer of gains from the calibrator sources to the target field. The calibration solutions were further refined via the process of self-calibration using the CASA package\cite{2007ASPC..376..127M}. For the imaging, W-projection was employed to account for the non-coplanar nature of the array\cite{2008ISTSP...2..647C,2005ASPC..347...86C}. For the weighting, we employed the  Briggs  scheme with a robust parameter of $0.0$, unless mentioned. 
 The 610~MHz image has a rms noise ($\sigma_{\rm{rms}}$) of 35~$\mu$Jy~beam$^{-1}$ and a resolution of 6.1\arcsec $\times$ 5.1\arcsec.  The 325~MHz image has a resolution of $10.5\arcsec\times 8.3\arcsec$ and $\sigma_{\rm{rms}}=87$~$\mu$Jy~beam$^{-1}$.

The cluster was also observed with  the Karl G. Jansky VLA, covering the  2--4~GHz S-band in B-array on 26~Feb, 2015 and in DnC-array on 7 Jan, 2016 (project SG0455). The  time on source was 1.2~hr for both observing runs. In addition, 1--2~GHz L-band observations (project 15A-270) were obtained in BnA-array  on 18~May, 2015 with a total time on source of 0.5~hr. These observations were reduced and calibrated with CASA following the process outlined in\cite{2016ApJ...817...98V}. RFI was automatically removed with the AOFlagger\cite{2010MNRAS.405..155O}. The calibration consisted of delay, bandpass,  and gain corrections. The channel dependent polarization leakage and angles were calibrated using the calibrators 3C147 and 3C138. The solutions were transferred to the target field and the solutions were further refined via the process of self-calibration. The imaging was carried out in the same way as the GMRT data. For imaging of the L-band data, we also used 1.4~GHz VLA D-array observations (project AC0696) to include emission on larger scales that are missed by the B-array observations. The reduction and calibration of the D-array observations have been described in\cite{2013ApJ...769..101V}. All images were corrected for the primary beam attenuation.

For making spectral index maps, we created images at 0.325 and 3.0~GHz. To correct for the different sampling densities in the uv-plane, uniform weighting was employed. We also placed inner uv-range cuts to image only common uv-ranges. The images were then convolved to the lowest resolution, given by the 325~MHz GMRT observations ($9.1\arcsec \times 6.5\arcsec$). The spectral index map for the area around~A (for the labeling see Fig.~\ref{fig:gmrt610}) is shown in the Supplementary Information (SI). We also computed a higher signal-to-noise spectral index map, including the 0.61 and 1.5~GHz observations. As before, the 0.61 and 1.5~GHz observations were imaged with uniform weighing and inner uv-range cuts. To compute the spectral index, we fitted power-law spectra through the four data points, ignoring any possible curvature. Pixels with values below $4\sigma_{\rm{rms}}$ were ``blanked''. The use of four frequencies decreases the noise on the derived spectral indices compared to only using  the  0.325 and 3.0~GHz images. The spectral index map and corresponding uncertainties, under the assumption that the spectra can be described by power-laws, are shown in the SI (a cutout around source~A is shown in Fig.~\ref{fig:spix}b).

The S-band polarization image is shown in Fig~\ref{fig:spix}c. This images has a resolution of $10\arcsec \times 8\arcsec$.
No vectors are plotted for pixels where the signal-to-noise ratio of the polarized intensity is $< 4$. Given the low Galactic Faraday Rotation Measure of $\approx -7$~rad~m$^{-1}$, we did not correct the polarization angles for Faraday Rotation \cite{2009ApJ...702.1230T}. The high-resolution S-band continuum map, showing the radio core associated with the AGN of source~A, is shown in the SI.

We computed the probability of a chance projection of the tailed radio galaxy with the relic attached to it. This was done by taking the ratio of the area covered by the relic and the cluster, with the cluster area given by $\pi*R_{500}^2$, and  $R_{500}\approx 1.4$~Mpc. This results in $p=0.002$. Taking the conservative approach, considering that the cluster contains two disturbed radio galaxies, the probability of a chance projection increases to $p=0.004$.

The cluster contains at least two more radio galaxies (B and C, see Fig.~\ref{fig:gmrt610}) to the south and southwest of the one described above. One of the radio galaxies (source~B) is embedded within the relic emission. Source~C is a distorted FR-I\cite{1974MNRAS.167P..31F} source.  The spectral index map and the 325~MHz image (SI and Fig.~\ref{fig:gmrt610}) reveal a hint of a connection between source's~C  southern lobe/tail and the southernmost part of the relic. The spectral index steepens along the lobes of source~C, as expected due to synchrotron and Inverse Compton losses. 
Given that the relic emission around sources~B and C is located further to south than the relic near source A, it is seems unlikely that the same shock is responsible for reviving the fossil plasma in all these regions. This would suggest that the ICM in the extreme southern outskirts of the 
 cluster is  disturbed by other shocks.  Our Chandra observations do show the presence of additional substructure in the general area around sources~B and C. However, given the low counts rates it is not possible to extract detailed information on the thermal properties of the ICM in this region.

 \subsubsection*{Chandra observations and data reduction.}

The cluster was observed with ACIS-I on the Chandra X-ray Observatory  for a total of 8 times between 2012 and 2015. This resulted in a total exposure time of 211~ks . The data were reduced with  the {\tt chav} package, following the processing described in\cite{2005ApJ...628..655V} and applying the CALDB 4.6.7 calibration files. This processing includes filtering of bad events by checking for periods of high background, corrections for the time dependence of the charge transfer inefficiency and gain, removal of readout artifacts, background subtraction, and exposure correction. For the background subtraction, we used standard blank sky background files. The images of the separate exposures were then combined into a single image, binning with a factor of 4 (i.e., 2\arcsec~pixel$^{-1}$). 


 \subsubsection*{Surface brightness profile fitting.}
\label{sec:proffit}
We fit the X-ray surface brightness profile with an updated version of Proffit\cite{2011A&A...526A..79E,2013MNRAS.433..812O}. We model the profile using an underlying broken power-law density model
\begin{equation}
  n(r)=%
  \begin{cases}
    C n_0 \left(\frac{r}{r_{\rm{edge}}} \right)^{a_2}  \mbox{ ,} &\text{$r < r_{\rm{edge}}$} \\
    \\
    n_0 \left(\frac{r}{r_{\rm{edge}}} \right)^{a_1}  \mbox{ ,}&\text{$r > r_{\rm{edge}}$}   \mbox{ .}
  \end{cases}
  \label{eq:densmodel}
\end{equation}
The subscripts 1 and 2 refer to the up and downstream regions, respectively. The parameter $r_{\rm{edge}}$ denotes the location of the jump, $n_0$ is the normalization constant, $C$ the shock compression factor, $a_1$ and $a_2$ are the slopes of the power-laws. This density model is then projected along the line of sight to obtain an X-ray surface brightness profile, assuming prolate spheroidal geometry within the sector. The emissivity is taken to be proportional to the density squared. 

In the case of a shock, the compression factor can be related to the shock Mach number
\begin{equation}
{\cal M}=\left[\frac{2 C}{\gamma + 1 - C(\gamma -1)}\right]^{1/2}   \mbox{ ,}
\label{eq:machne}
\end{equation}
\noindent where $\gamma$ is the adiabatic index of the gas. We assume $\gamma=5/3$ for the thermal plasma.

We fit the surface brightness profile  in an elliptical sector that crosses the radio relic (see the SI), using the model given in Eq.~\ref{eq:densmodel}.  The opening angles ($210\degr - 250\degr$) were chosen after visual inspection of the Chandra image (see the SI), which indicates a possible surface brightness edge at the relic location and extending $\sim 10\degr$ further to the west. For the radial binning of photons, we require a SNR of 5 per bin. We exclude the regions affected by point sources during the fitting. The observed profile and most best fitting model is shown in Fig.~\ref{fig:profile}. We find a break in the X-ray surface brightness profile at the location of the radio relic. The resulting density jump is small $C=1.31^{+0.10}_{-0.09}$ ($\mathcal{M}=1.2$), with a 90\% confidence upper limit of $C<1.56$, indicating that the Mach number is low $\mathcal{M}<1.4$. The 90\% lower confidence limit is $C > 1.16$. The location of the jump is indicated in Fig.~\ref{fig:spix}b. 

We also extracted a profile in a spherical sector (see the SI). The resulting fitted compression factor is slightly lower with $C\approx1.2$, but consistent with the previous results. In addition, we checked whether the results changed for a smaller opening angle, corresponding to the relic's visible extent in the 610 and 325~MHz images (i.e., an opening angle of $210\degr - 240\degr$). This did not result in significant changes for the values of the compression factor or discontinuity location. We also slightly varied the placement of the sector, again obtaining consistent results. Therefore we conclude that the presence of a jump does not depend on the precise sector placement and shape. For comparison, in the SI we show a model with a density jump of $C=2.3$ ($\mathcal{M}=2.0$) in the elliptical sector, which does not provide a good match to the data. 

A small ``bump'' is visible in the X-ray surface brightness profile at a radial distance of 4.5\arcmin - 5.0\arcmin~(Fig.~\ref{fig:profile}).  The nature of this bump is unclear. Ignoring this bump increases the best-fitting compression factor and the 90\% confidence upper limit to $C<2.0$ ($\mathcal{M}< 1.7$). 

From the X-ray surface brightness profile we conclude that there is evidence for a deviation from hydrostatic equilibrium as expected if a shock were present. The location of the discontinuity coincides with the location of the relic, and also agrees with the location found by visual inspection of the Chandra image. The modeling suggests that the underlying density jump must be rather low ($C<2.0$, $\mathcal{M}< 1.7$).  Some caution is required in interpreting the derived Mach number and upper limits, since it depends on the assumptions made in the modeling. Given the extra substructure in this region, as seen in the Chandra image, some of the assumptions might be incorrect. The unknown projection effects and shock geometry typically result in underestimation of the Mach number and our derived values should therefore be considered as lower limits.

Another way to constrain the Mach number, and rule out the presence of a cold front, is to determine the temperatures on both sides of the discontinuity (in a small enough region not be affected by other cluster substructure). An advantage of this method is that temperature measurements are less affected by the unknown geometry. However, the count rate in the region south of the discontinuity is very low, and therefore we are not able to obtain useful constraints on the pre-shock X-ray gas temperature.

\subsubsection*{SOAR spectroscopy and data reduction.}

The optical spectrum of the AGN host galaxy that ``fuels'' the relic (RA 08$^{\rm{h}}$42$^{\rm{m}}$13$^{\rm{s}}$.73; DEC $-$17\degr31\arcmin12\arcsec.1) was acquired using the Goodman Spectrograph on the Southern Astrophysical Research Telescope (SOAR) telescope. The observations were conducted on November 19, 2015, as part of program SO2015B-020. The basic setup included the 600~l~mm$^{\rm{-1}}$ grating (blue setting), and a 1$\farcs$03 slit. The wavelength range of the observed spectrum was 3500-5500\,{\AA}, with a resolving power of $R\sim1500$ and S/N$\sim10$~pixel$^{-1}$ at 4500\,{\AA}. Calibration frames included Cu and HgAr arc-lamp exposures, quartz-lamp flat-fields, and bias frames. The data reduction, including spectral extraction and wavelength calibration, were performed using standard IRAF routines (\url{http://iraf.noao.edu}).

A SOAR optical spectrum of the AGN host galaxy is shown in the SI. From the spectrum we determine a redshift of $z=0.164\pm0.001$, consistent with the galaxy being a cluster member.

\subsubsection*{Subaru and Keck observations.}
\label{sec:SubaruKeck}

We carried out deep imaging observations of the Abell~3411-3412 system with Subaru SuprimeCam\cite{2002PASJ...54..833M} on 2014 February 24 in $g$, $r$ and $i$ with integrations of 720\,s, 2880\,s, and 720\,s, respectively (P.I. D. Wittman).
We took 4 exposures for $g$ and $i$, and 8 exposures for $r$.
We rotated the field between each exposure (30\degr~for $g$ and $i$, and 15\degr~for $r$) in order to distribute the bleeding trails and diffraction spikes from bright stars azimuthally and later removed them by median-stacking exposures. 
This scheme enables us to maximize the number of detected galaxies, especially considering the number of stars at A3411-3412's low galactic latitude (+15\degr).
The average seeing for the images was $\sim0.\arcsec85$.
The details of the Subaru data reduction are similar to those presented in\cite{2015ApJ...802...46J}.

The primary objective for the spectroscopic survey was to maximize the number of cluster member spectroscopic redshifts.
Since the SuprimeCam imaging was unavailable at the time of the spectroscopic survey planning, we took images with the Isaac Newton Telescope Wide Field Camera (INT/WFC),  on 2013 October 31 in $g$ and $i$ bands,  to determine the approximate red sequence of the cluster  and preferentially selected those galaxies followed by potential blue cloud galaxies. 
These INT/WFC data were reduced with our in-house PYTHON-based pipeline\cite{2014MNRAS.438.1377S,2015MNRAS.450..630S}
Briefly, the sky flats for each filter were median-combined to obtain a `master-flat'. A `master-bias' was obtained by median-combining biases. The individual exposures were then bias-subtracted and sky-flattened. Astrometric solutions were obtained by using SCAMP\cite{2006ASPC..351..112B}, and images were zero-point calibrated with the fourth United States Naval Observatory (USNO) CCD Astrograph Catalog (UCAC4)\cite{2013AJ....145...44Z}, before being stacked/combined using SWARP\cite{2002ASPC..281..228B}. The difficulty of star-galaxy separation is compounded by the 1.5$\arcsec$-2$\arcsec$ seeing of the INT/WFC imaging, which results in many blended pairs of stars passing morphological cuts designed to eliminate point sources.

We observed the Abell~3411-3412 system with the DEIMOS\cite{2003SPIE.4841.1657F} instrument on the Keck II 10\,m telescope on 2013, December~3 and~4. 
Observations were taken using 1\arcsec\ wide slits with the 1200\,line\,mm$^{-1}$ grating, resulting in a pixel scale of $0.33$\,\AA\,pixel$^{-1}$ and a resolution of $\sim1$\,\AA\ (50\,km\,s$^{-1}$).
The grating was tilted to a central wavelength of 6650\,\AA, resulting in a typical wavelength coverage of 5350\,\AA\ to 7950\,\AA, which encompasses the spectral features H$\beta$, [\ion{O}{3}] 4960 \& 5008, \ion{Mg}{1} (b), \ion{Fe}{1},  \ion{Na}{1} (D), [\ion{O}{1}], H$\alpha$, and the [\ion{N}{2}] doublet for galaxies near the cluster redshift.
The actual wavelength coverage may be shifted by $\sim\pm410$~\AA\, depending on where the slit is located along the width of the slitmask.
We observed a total of four slit masks with approximately 120 slits per mask.
For each mask we took three 900\,s exposures.
The data reduction followed the same procedure outlined in detail in\cite{2015MNRAS.450..630S,2015ApJ...805..143D}.

We obtained 484 spectra with DEIMOS. Of these, we obtained reliable redshifts for 447 objects (92\%), leaving 37 spectra that were either too noisy, had ambiguous redshift solutions (e.g., those with a single emission line), or failed to reduce properly.  Of the 447 reliable redshifts, 221 (49\%) are stars, indicating the difficulty of star/galaxy discrimination with low-resolution imaging at low Galactic latitude.  Adopting the quality rating system of\cite{2013ApJS..208....5N}, in which only galaxies with secure redshifts ($Q>3$) are considered high-quality, the removal of the stars results in 226 high quality DEIMOS galaxy spectra.  Of these, 174 (77\%) fall within $0.148 \leq z \leq 0.176$ (see the SI, which is  $z_\mathrm{cluster}\pm 3\times\sigma$, where $z_\mathrm{cluster}=0.162$ and $\sigma$ is the approximate velocity dispersion ($1000\,\mathrm{km}\,\mathrm{s}^{-1}$). The remaining 52 high-quality spectra consist of one foreground and 51 background galaxies, comprising 0.4\% and 23\% of the high-quality spectra respectively.
The subcluster identification and the dynamical modeling are described in the SI.

\end{methods}

\clearpage
\newpage
\section{Supplementary Information}

\subsection{Additional radio images.}

\begin{figure*}[h!]
    \begin{center}

\includegraphics[angle = 180, trim =0cm 0cm 0cm 0cm,width=0.49\textwidth]{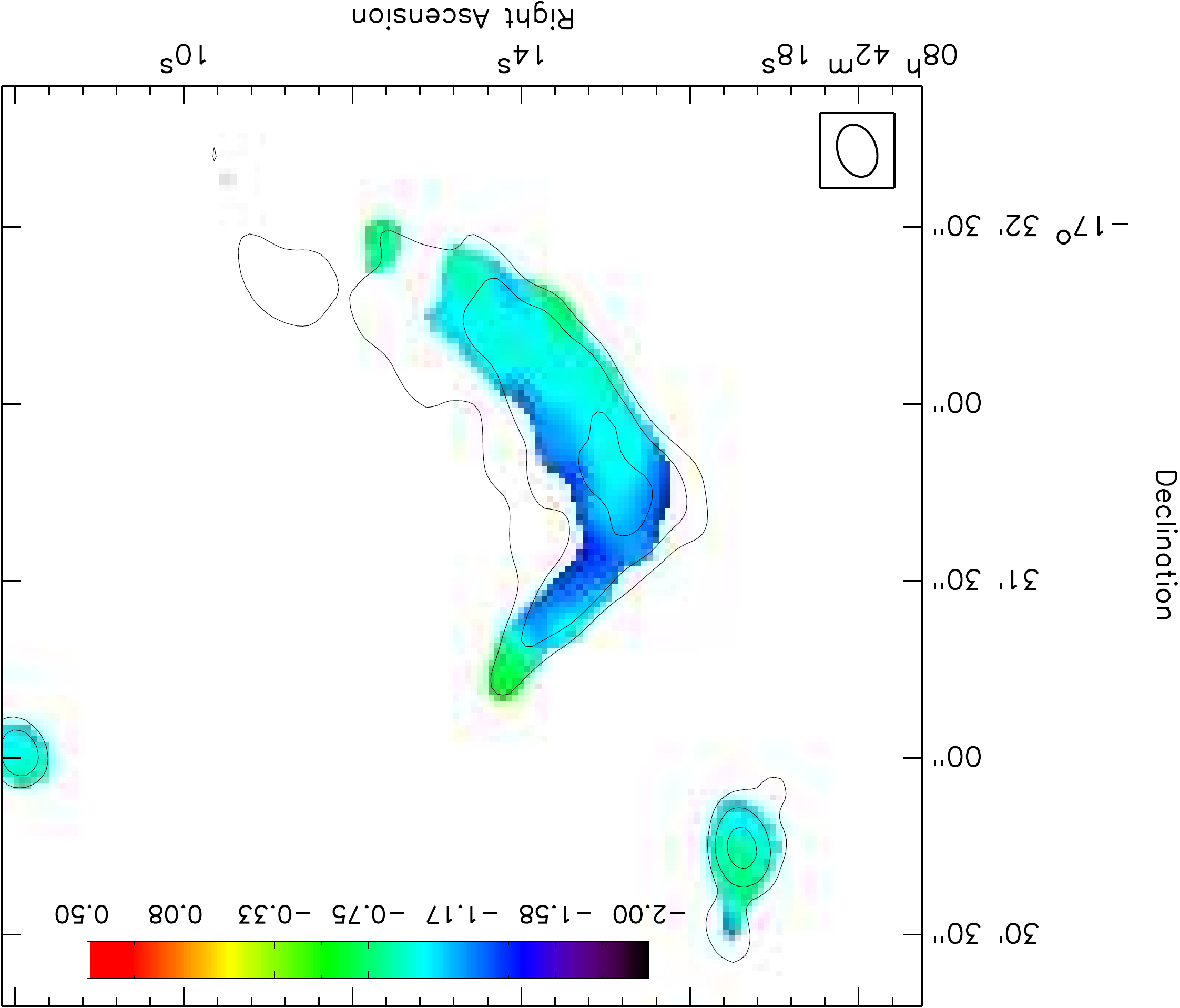}  
\includegraphics[angle = 180, trim =0cm 0cm 0cm 0cm,width=0.5\textwidth]{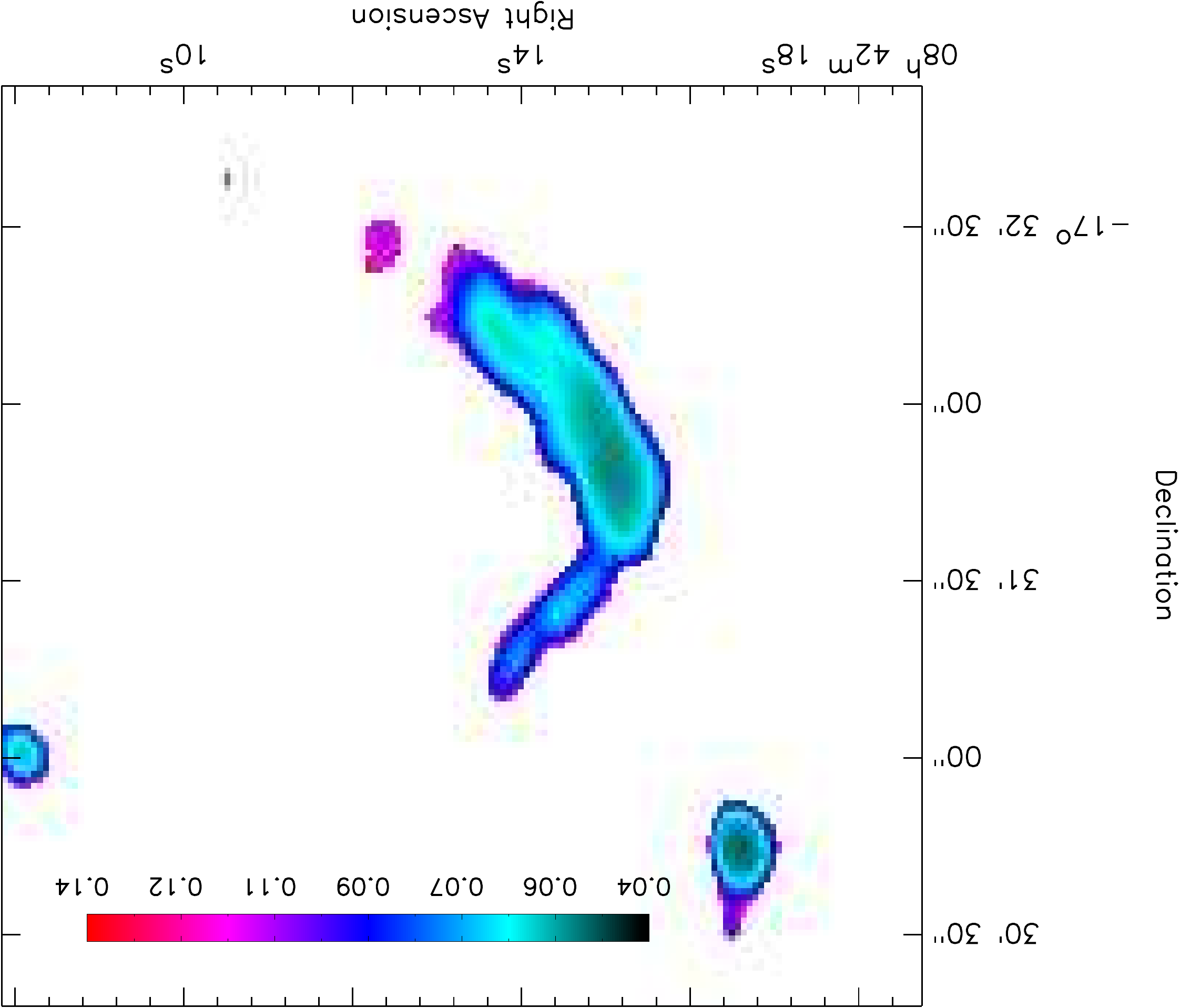}
       \end{center}
       \caption{Left: Spectral index map at a resolution of $9.1\arcsec \times 6.5\arcsec$ between 0.325 and 3.0~GHz. The contours  are from the 325~MHz image and are drawn at levels of ${[1,2,4,\ldots]} \times 6\sigma_{\rm{rms}}$. Pixels with values below $5\sigma_{\rm{rms}}$ were blanked. The beam size is indicated in the bottom left corner. Right: Spectral index uncertainty map ($1\sigma$) corresponding to the figure shown in the left panel.}
      \label{fig:spixsmall}
 \end{figure*}

 \begin{figure*}[h!]
    \begin{center}
      \includegraphics[angle = 180, trim =0cm 0cm 0cm 0cm,width=0.8\textwidth]{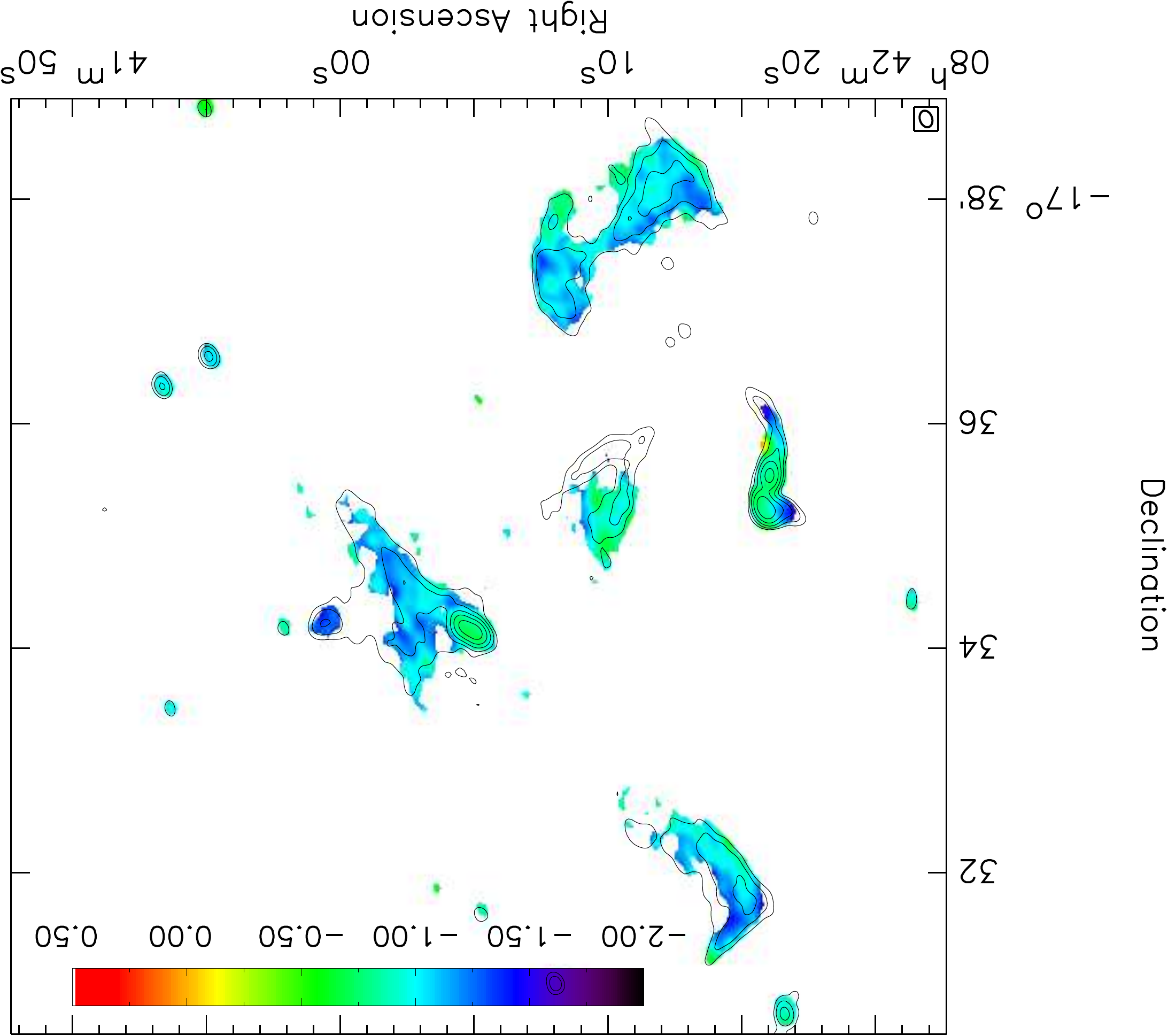}
       \end{center}
       \caption{Spectral index map using observations at 0.325, 0.61, 1.5, and 3.0~GHz fitting a single power-law through the flux density measurements. The spectral index image has a resolution of $9.1\arcsec \times 6.5\arcsec$. Pixels with values below $3\sigma_{\rm{rms}}$ were blanked. Contour levels are from the 325~MHz image and are drawn at levels of ${[1,2,4,\ldots]} \times 6\sigma_{\rm{rms}}$. The beam size is indicated in the bottom left corner.}
            \label{fig:spixand325}
 \end{figure*}

 \begin{figure*}[h!]
    \begin{center}
      \includegraphics[angle = 180, trim =0cm 0cm 0cm 0cm,width=0.8\textwidth]{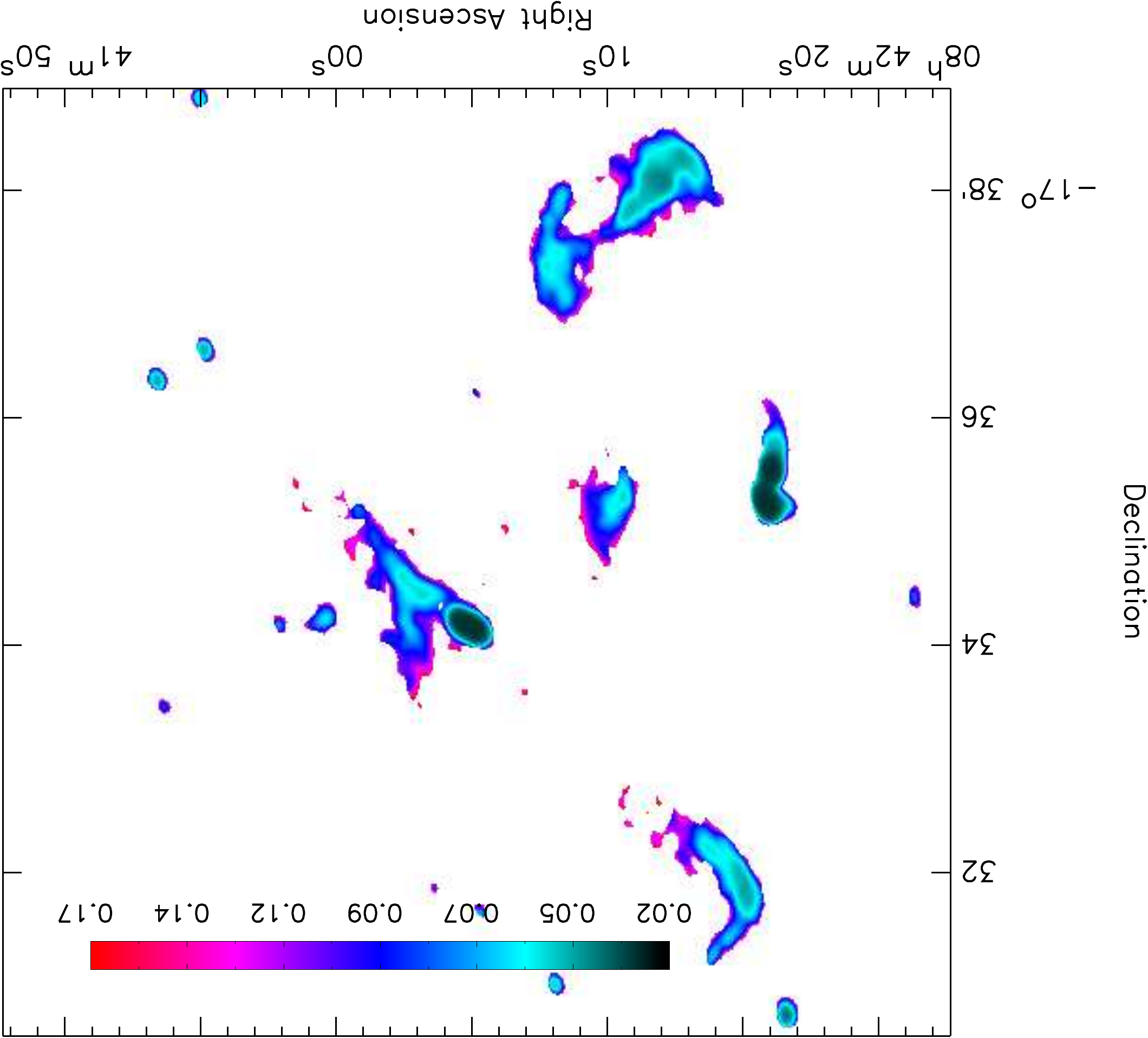}      
       \end{center}
       \caption{Spectral index uncertainty ($1\sigma$) map corresponding to Supplementary Fig.~\ref{fig:spixand325}.}
            \label{fig:spixanderr4freq}
 \end{figure*}

 \begin{figure}[h!]
    \begin{center}
      \includegraphics[angle = 180, trim =0cm 0cm 0cm 0cm,width=0.7\textwidth]{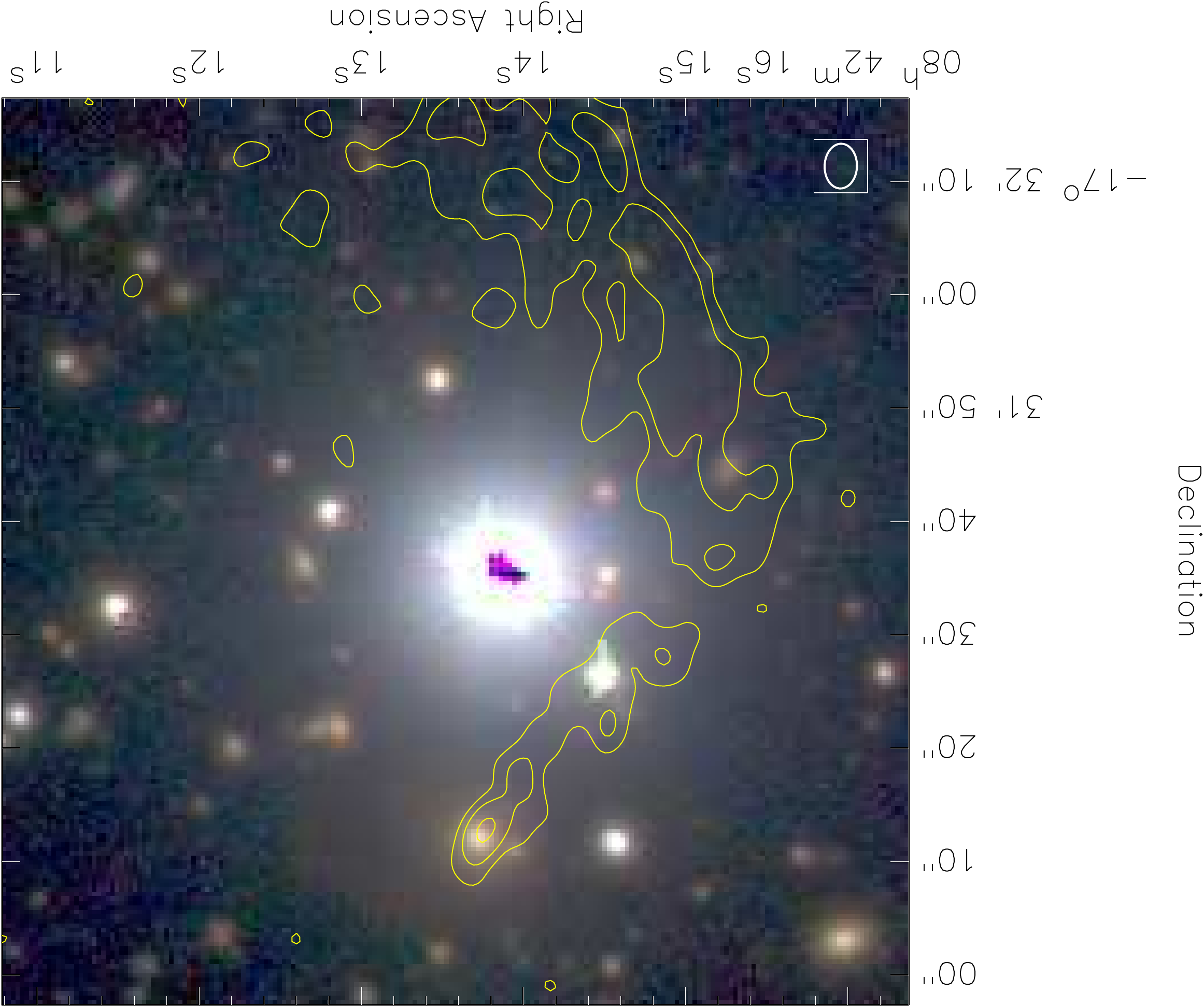}
      \label{fig:sbandhighres} 
       \end{center}
      \caption{Subaru \emph{gri} color image with 3~GHz VLA contours overlaid showing the radio tail originating from the host galaxy located at the upper part of the image. The bright saturated object close to the center of the image is a foreground star.   Radio contours are drawn at $[1, 2, 4,\ldots]\times$~$2.5\sigma_{\rm{rms}}$, with $\sigma_{\rm{rms}} = 6$~$\mu$Jy~beam$^{-1}$. The resolution of the radio image is $3.9\arcsec\times2.9\arcsec$. The beam size is indicated in the bottom left corner.}
\end{figure} 
\clearpage

\section{Images for the X-ray surface brightness profile modeling.}

 \begin{figure*}[h!]
    \begin{center}
      \includegraphics[angle = 0, trim =0cm 0cm 0cm 0cm,width=0.43\textwidth]{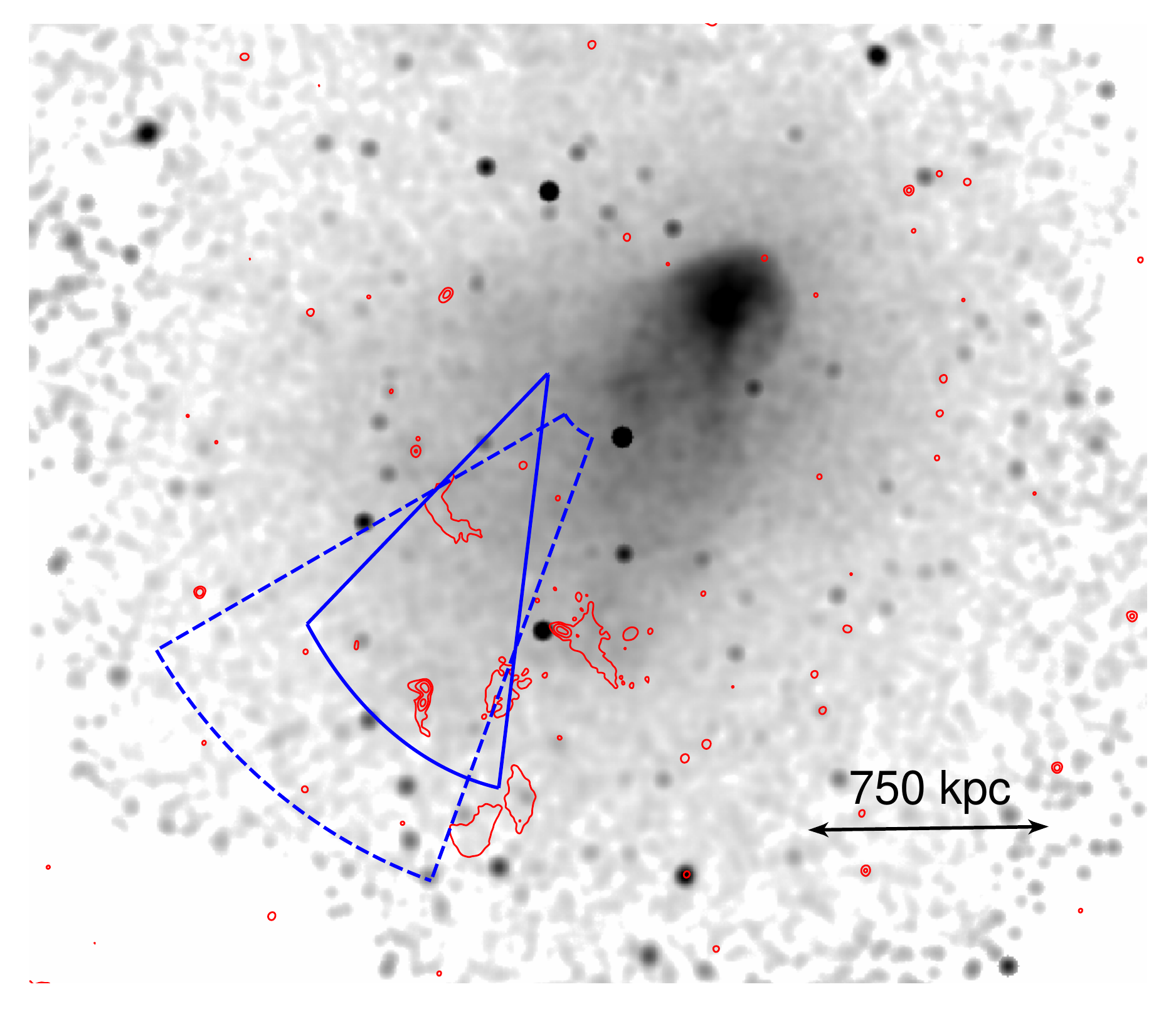}
           \includegraphics[angle = 0, trim =0cm 0cm 0cm 0cm,width=0.516\textwidth]{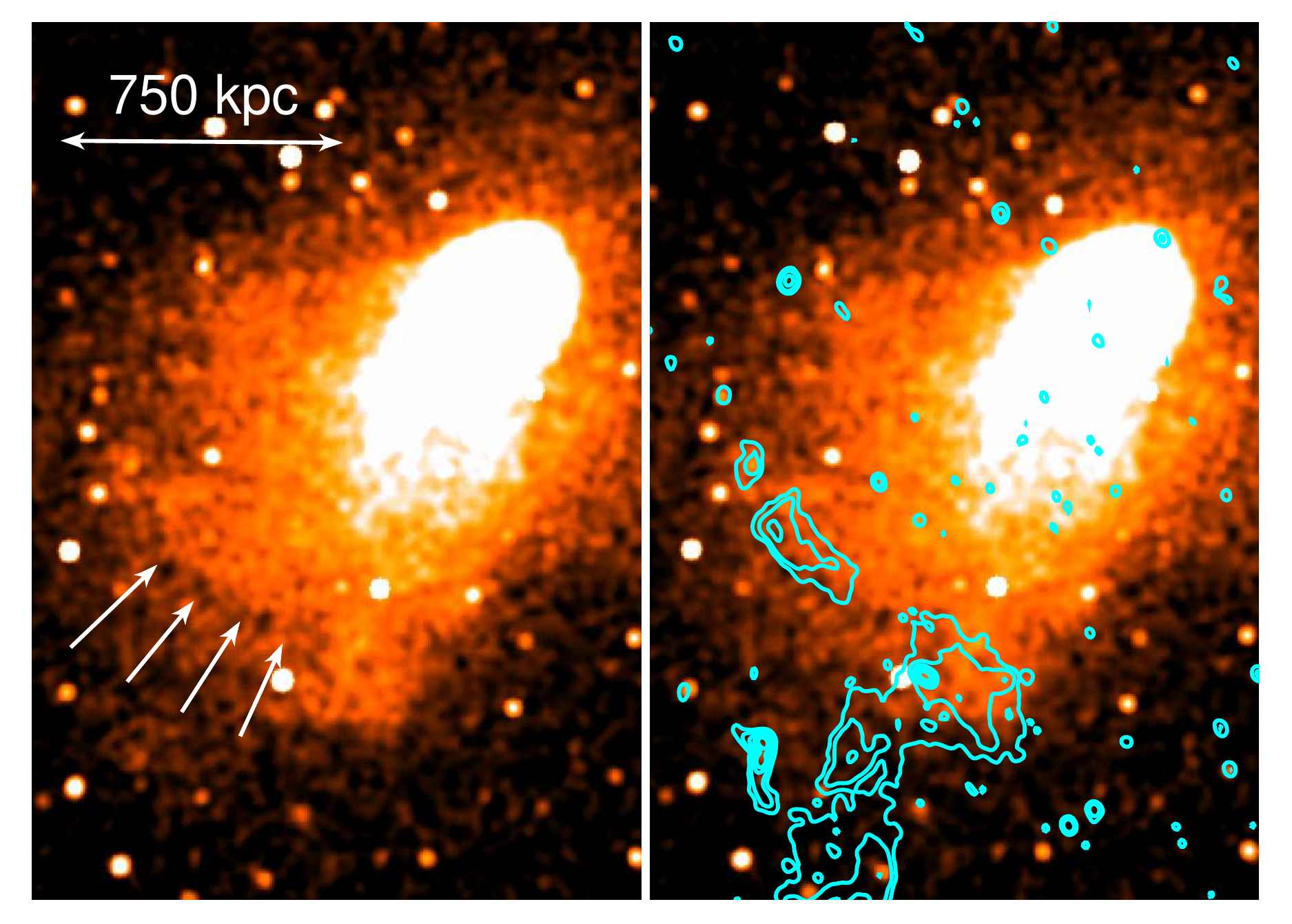}
       \end{center}\vspace{0mm}
       \caption{Left: Chandra 0.5--2.0~keV image, smoothed with a Gaussian of $8\arcsec$, indicating the position of the elliptical sector (blue) in which we modeled the X-ray surface brightness profile. Radio contours  from the GMRT 610~MHz image are shown in red. Contour levels are drawn at $[1,2,4,\ldots] \times 200$~$\mu$Jy~beam$^{-1}$. The alternative sector for modeling and extracting the X-ray surface brightness profile is shown with dashed lines. Right:  Chandra 0.5--2.0~keV image indicating the location of the surface brightness edge with arrows. In right sub-panel the same image is shown with contours from the GMRT 325~MHz image overlaid.\vspace{5mm}}
              \label{fig:wedge}
 \end{figure*}

\begin{figure*}[h!]
    \begin{center}
                \includegraphics[angle = 0, trim =0cm 0cm 0cm 0cm,width=0.49\textwidth]{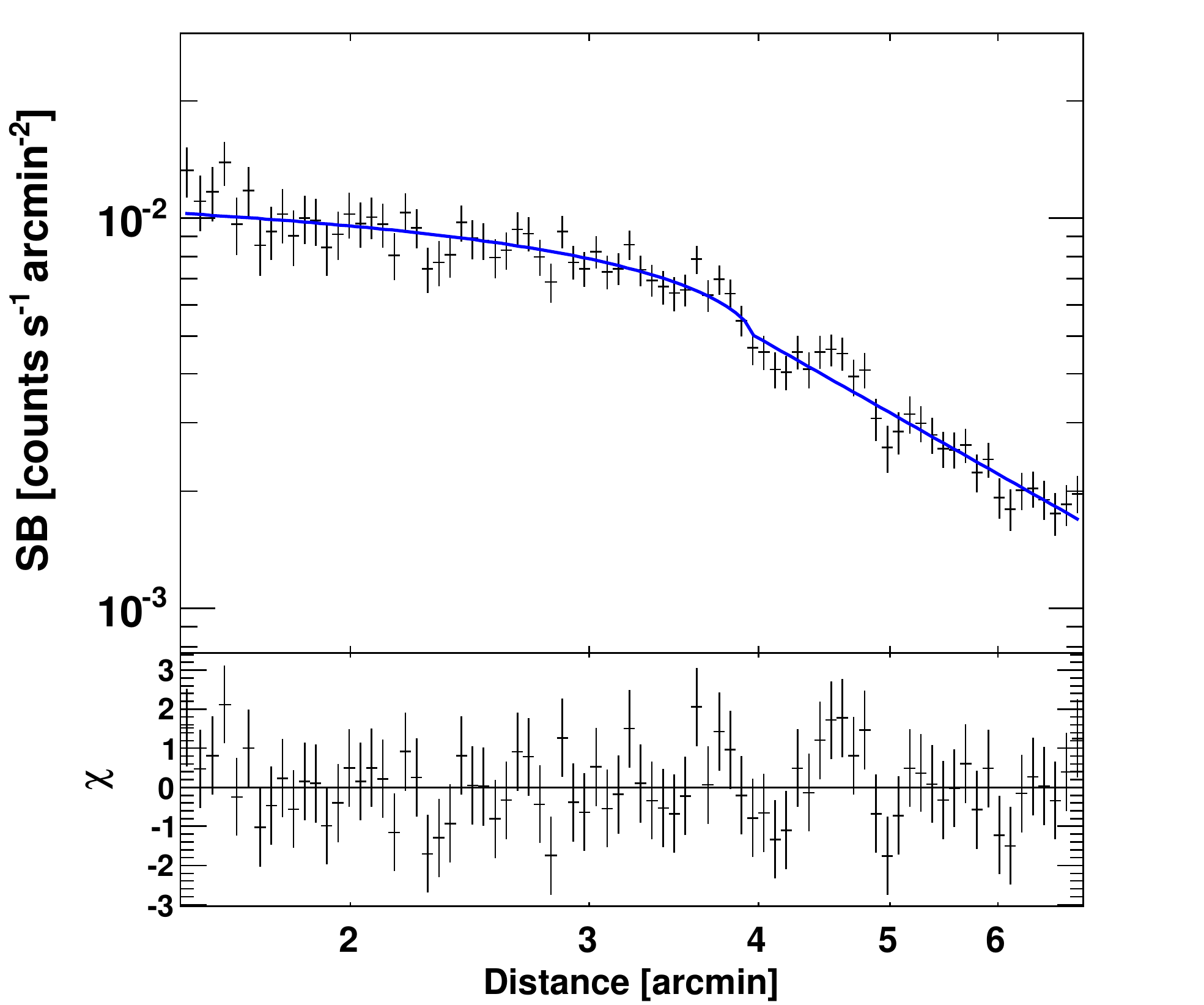}
      \includegraphics[angle = 0, trim =0cm 0cm 0cm 0cm,width=0.49\textwidth]{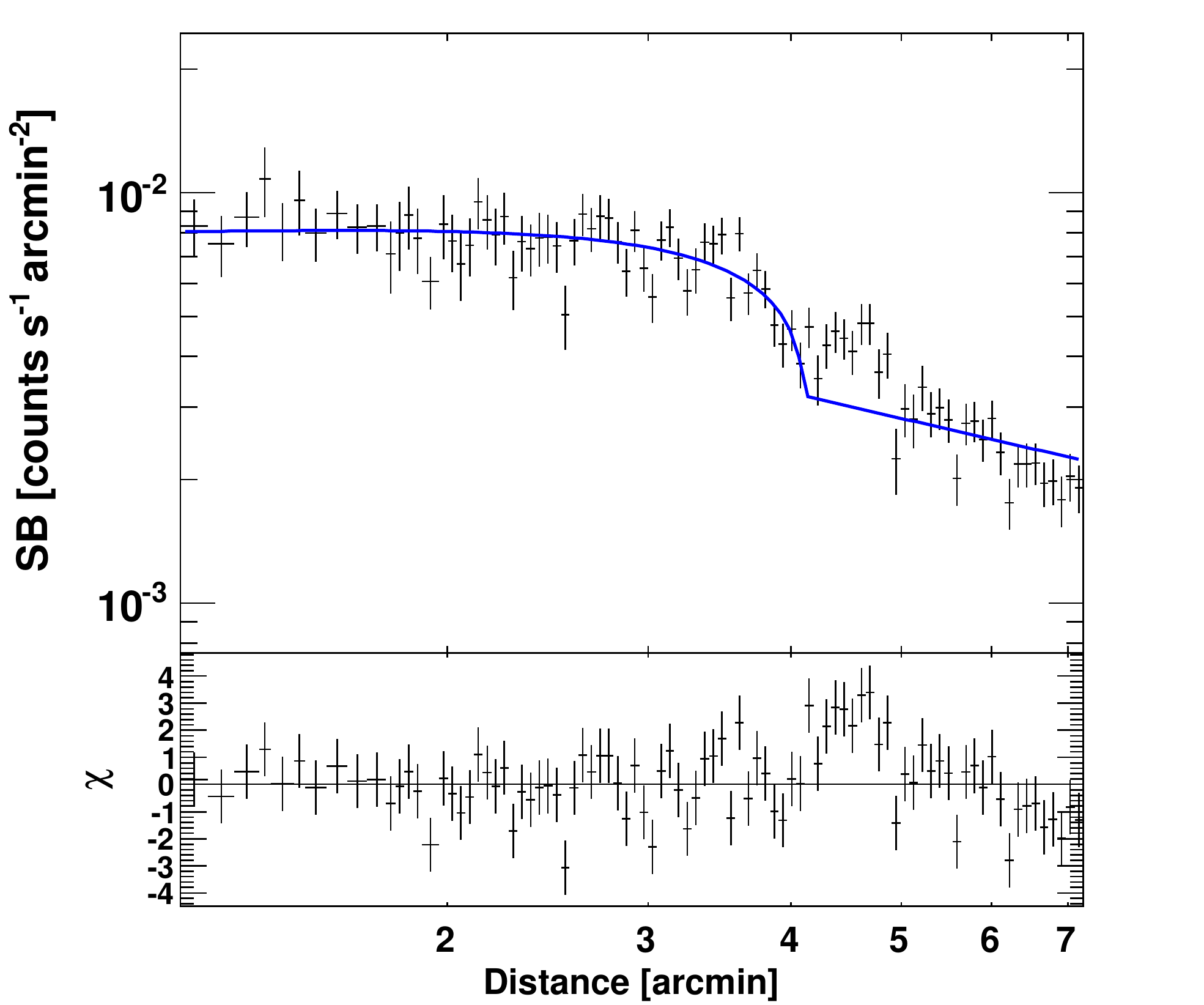}
       \end{center}\vspace{0mm}
       \caption{Left: Chandra 0.5--2.0~keV surface brightness profile across the radio relic in the spherical sector shown in Supplementary Fig.~\ref{fig:wedge} (left panel).  Uncertainties on the data points are plotted at the $1\sigma$ level. The blue line corresponds to the best-fitting double power-law model which has $C=1.2$. Right: The same figure as in main article (Fig.~4), but where we fixed the density jump to $C=2.3$ ($\mathcal{M}=2.0$) during the fit. }
       \label{fig:profileSUP}
 \end{figure*}

 \subsection{SOAR optical spectrum}

 \begin{figure}[h!]
    \begin{center}
      \includegraphics[angle = 0, trim =0cm 0cm 0cm 0cm,width=0.5\textwidth]{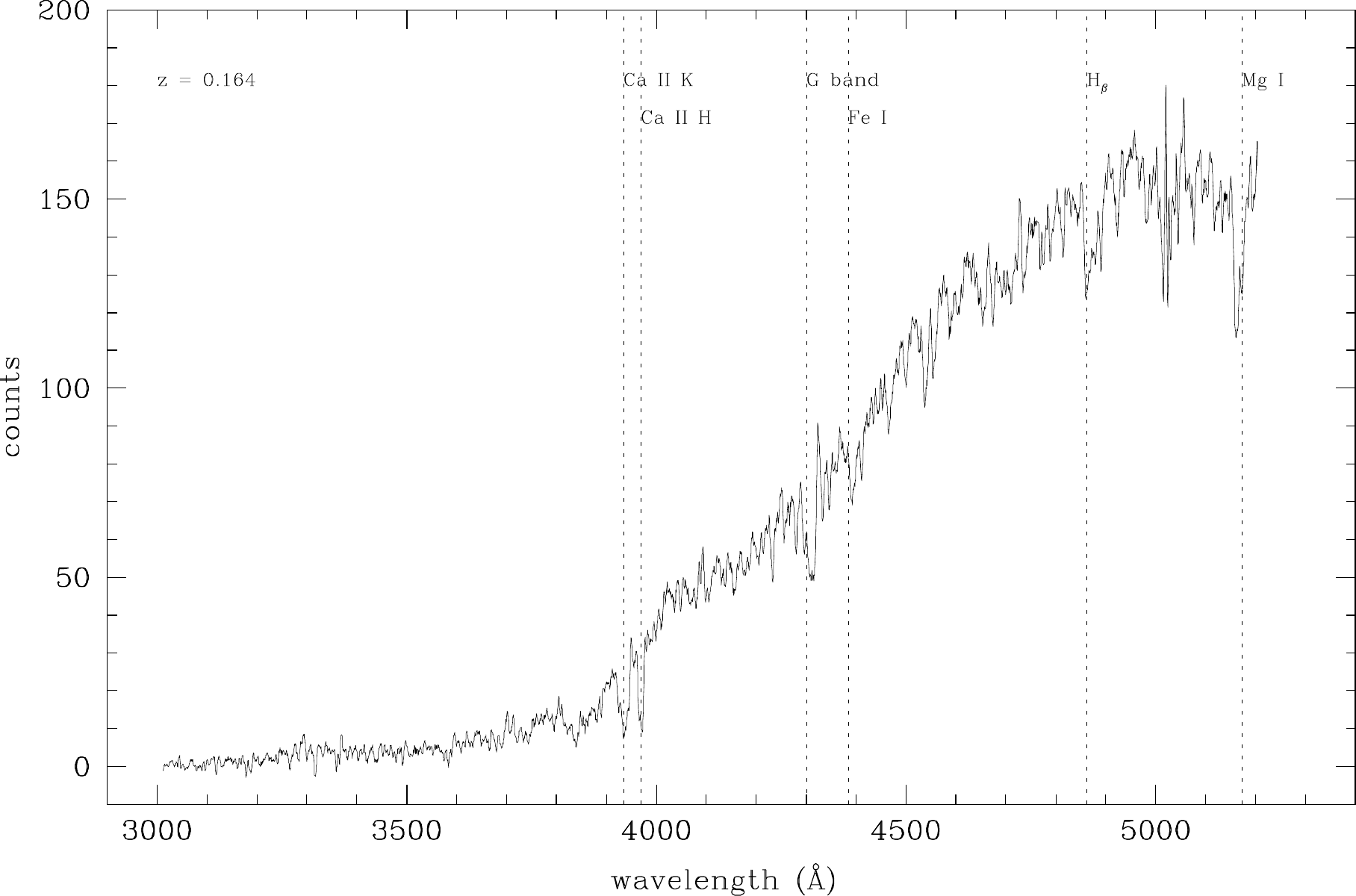}
       \end{center}
       \caption{SOAR spectrum of the AGN host galaxy ($z=0.164\pm0.001$). The positions of some spectral features are indicated.}
       \label{fig:opticalspec}
 \end{figure}

 \subsection{Radio spectral re-acceleration modeling.}

We this section we evaluate one possible re-acceleration model to investigate the spectral index distribution across the relic. We first created a radio profile across the width of the relic, using the regions indicated in Supplementary Fig.~\ref{fig:radioregion}. The resulting radio profiles are displayed in Supplementary Fig.~\ref{fig:profileKang}.  

For modeling the spectral change during re-acceleration and the radio profiles in the shock downstream region, we follow the method outlined by\cite{2015ApJ...809..186K} for re-acceleration. We assume that a spherical shock sweeps through a magnetized gas cloud containing  fossil relativistic electrons, while propagating through a density gradient of $r^{-2}$ in the cluster outskirts. The simulated profiles are convolved to a spatial resolution of 20~kpc.  The post-shock temperature is set to $6.4^{+0.6}_{-0.5}$~keV, which was determined from the Chandra data in the region north of the relic. The radius of the shock is assumed to be 1.0~Mpc. The geometry of the cloud is the same as described in\cite{2015ApJ...809..186K}.  The pre-shock magnetic field is 1.5~$\mu$Gauss, and the magnetic field strength across the shock transition is assumed to increase due to the compression of two perpendicular components. Note that the shock speed and the post-shock magnetic field strength vary in time as the shock slows in the cluster outskirts.

 \begin{figure*}[h!]
    \begin{center}
      \includegraphics[angle = 0, trim =0cm 0cm 0cm 0cm,width=1.0\textwidth]{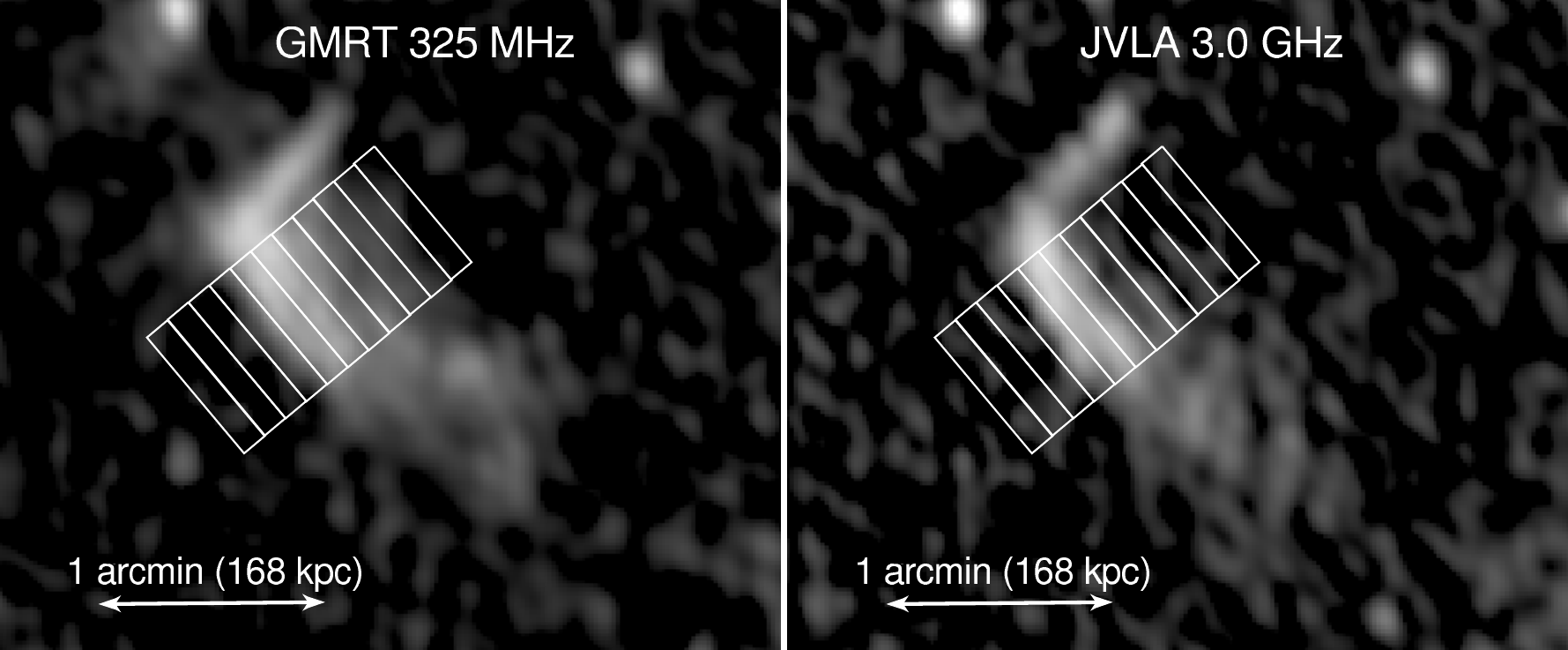}
       \end{center}\vspace{0mm}
       \caption{Regions where the radio profiles across the relic were extracted displayed on the GMRT 325~MHz (left) and VLA 3.0~GHz image (right). The width of each rectangular box is 20~kpc. The profiles are shown in Supplementary Fig.~\ref{fig:profileKang}.}
            \label{fig:radioregion}
 \end{figure*}

The fossil electrons are assumed to have the energy spectrum, $N(\gamma_e)\propto \gamma_e^{-2.5} \exp[-(\gamma_e/\gamma_{e,c})^2]$, where the cutoff Lorentz factor is $\gamma_{e,c} = 1.8 \times 10^4$.  The slope of $s=2.5$ is a typical injection index for AGN \cite{2008MNRAS.390..595M},  which corresponds to a radio spectral index of $\alpha = -0.75$ ($\alpha = (s-1)/2$). The cutoff ($\gamma_{e,c} =  1.8 \times 10^4$) at the end of the tail, just before re-acceleration, is set to match the $\alpha=-1.3$ between 0.325 and 3.0~GHz. The ``extension angle" is $\psi\approx 12^{\degr}$\cite{2015ApJ...809..186K}.

The best matching model has a sonic Mach number of $\mathcal{M}_s = 1.9$ at the time shown in Supplementary Fig.~\ref{fig:profileKang}, where we can reproduce the spectral change from $\alpha=-1.3$ just before re-acceleration to $\alpha=-0.9$ at the relic.  The post-shock magnetic field has a value of 2.9~$\mu$Gauss. We note that the observed profiles  are wider than the simulated ones and that the Mach number is somewhat higher than the upper limit determined from the X-ray observations. In our modeling, we assumed the relic is seen edge-on with respect to the shock normal. If this is not the case, the observed profiles would widen. Our Mach number estimates are also somewhat uncertain due to the assumptions that were made when fitting the X-ray surface brightness profile. We also evaluated a model with a flatter injection index of $s=2.2$ ($\alpha=-0.6$). In that case, we find a best matching  sonic Mach number of $\mathcal{M}_s = 1.7$. Given the uncertainties in the geometry, simplifying assumptions made in the modeling, and unknown distribution of fossil plasma before re-acceleration, we consider the match with the data reasonable.

It is expected that for a tailed radio galaxy, the shape of the energy spectrum of the fossil electrons changes with distance from the nucleus. Thus the spectral index along the shock front, towards the west, is expected to steepen. However, the observed spectral index along the relic's extent is rather uniform. We therefore extended our model to investigate what spectral index trend is to be expected, taking into account the predicted spectral steepening along the original tail of fossil plasma.

We first estimate the amount of spectral steepening that is expected along the 150~kpc extent of the relic in the spectral index map. Assuming that the distance between the core and the end of the tail (where $\alpha= -1.3$) is about 90~kpc, the mean jet velocity is estimated to be $1.2 \times 10^3$~km~s$^{-1}$ (for $s=2.5$, $\gamma_{e,c}  = 1.8 \times 10^4$, and a cooling time of $7.3 \times10{^7}$~yr). The distance from the NE corner to the SW corner of the relic is about 150~kpc. 
The cooling or advection time of the jet is estimated to be $1.2\times10^8$ yr over this 150~kpc. In that case, the cutoff Lorentz factor should decrease from $\gamma_{e,c}=1.8\times10^4$ to $\gamma_{e,c} = 1.2\times10^4$.  This leads to spectral ageing from $\alpha = -1.3$ to $\alpha = -1.6$ along the
relic's length. When then rerun our model, changing the input $\gamma_{e,c}$. The resulting spectral profiles after re-acceleration are shown in Supplementary Fig.~\ref{fig:profilesteeperfossiltrend}. At the relic's NE corner with $\alpha = -1.3$, re-acceleration changes the spectral index to $-0.9$, the result we obtained before. At the SW corner with $\alpha = -1.6$,  re-acceleration changes the spectral index back to $\alpha=-1.0$. So we can see that the re-acceleration leads to a relatively uniform spectral index along the relic's extent, even though the original fossil distribution shows considerable change. While this model does explain the relatively uniform spectral index along this particular relic, it is worth noting that it might have trouble explaining the constant spectral indices for Mpc-size relics along their lengths (unless the shock Mach number is higher or the spectral index distribution of the fossil plasma does not show extremely large variations). In addition, there could be other challenges for the ``DSA-type'' re-acceleration model described in this section, see \cite{2014MNRAS.437.2291V,2015MNRAS.451.2198V}.

\begin{figure*}[h!]
    \begin{center}
      \includegraphics[angle = 180, trim =0cm 0cm 0cm 0cm,width=0.75\textwidth]{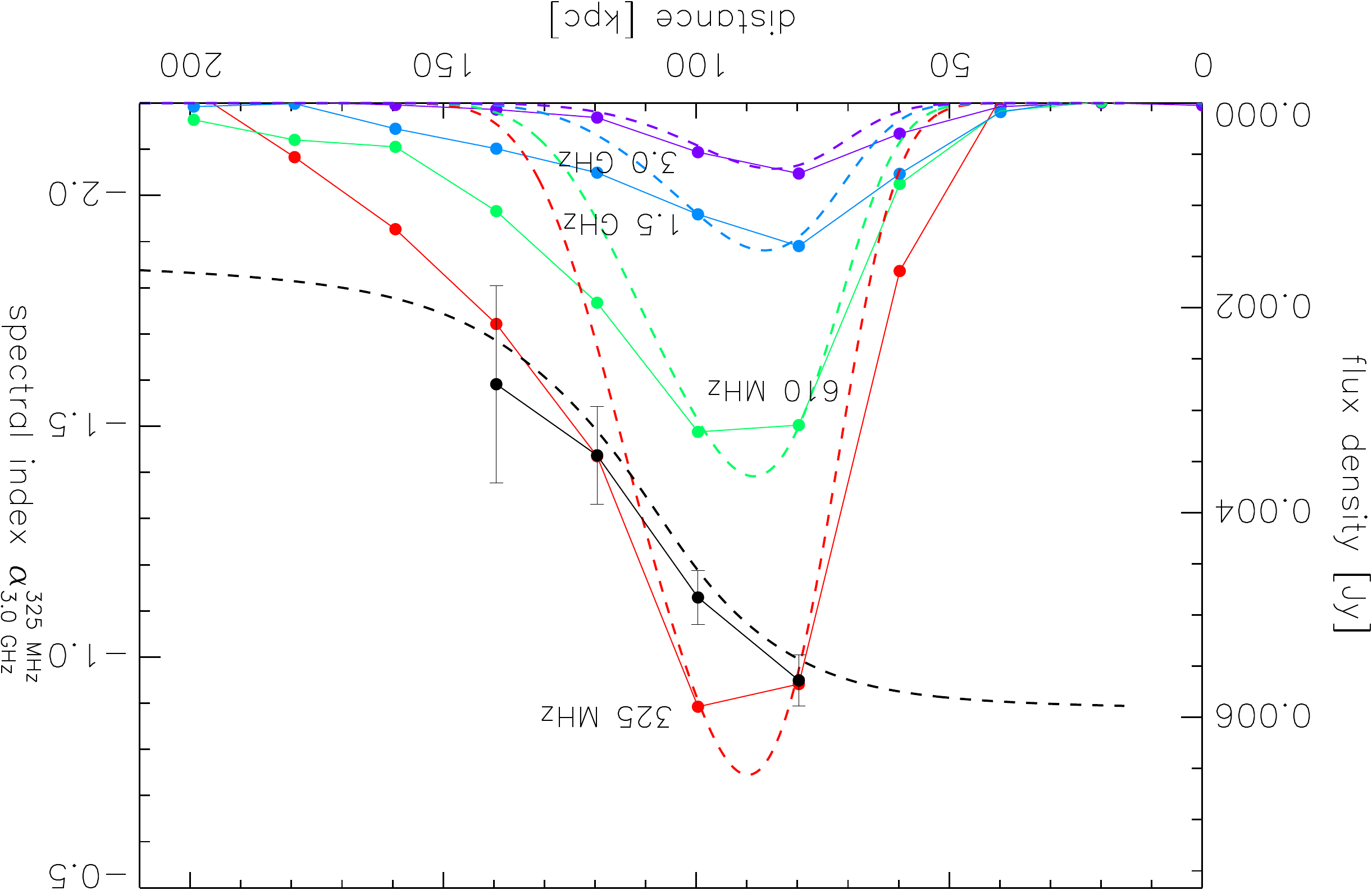}     
            \end{center}
       \caption{Measured radio flux density profiles (red: 325~MHz, green: 610~MHz, blue: 1.5~GHz, and purple: 3.0~GHz) across the width of the relic (see Supplementary Fig.~\ref{fig:radioregion}).  The spectral index ($\alpha^{\rm{325~MHz}}_{\rm{3.0~GHz}}$) profile is shown in black. The data points are connected by solid lines. 
       Uncertainties are plotted at the $1\sigma$ level. Modeled profiles from re-accelerated fossil plasma are overlaid with dashed lines with corresponding colors. }
       \label{fig:profileKang}
 \end{figure*}

\begin{figure*}[h!]
    \begin{center}
      \includegraphics[angle = 0, trim =0cm 0cm 0cm 0cm,width=0.75\textwidth]{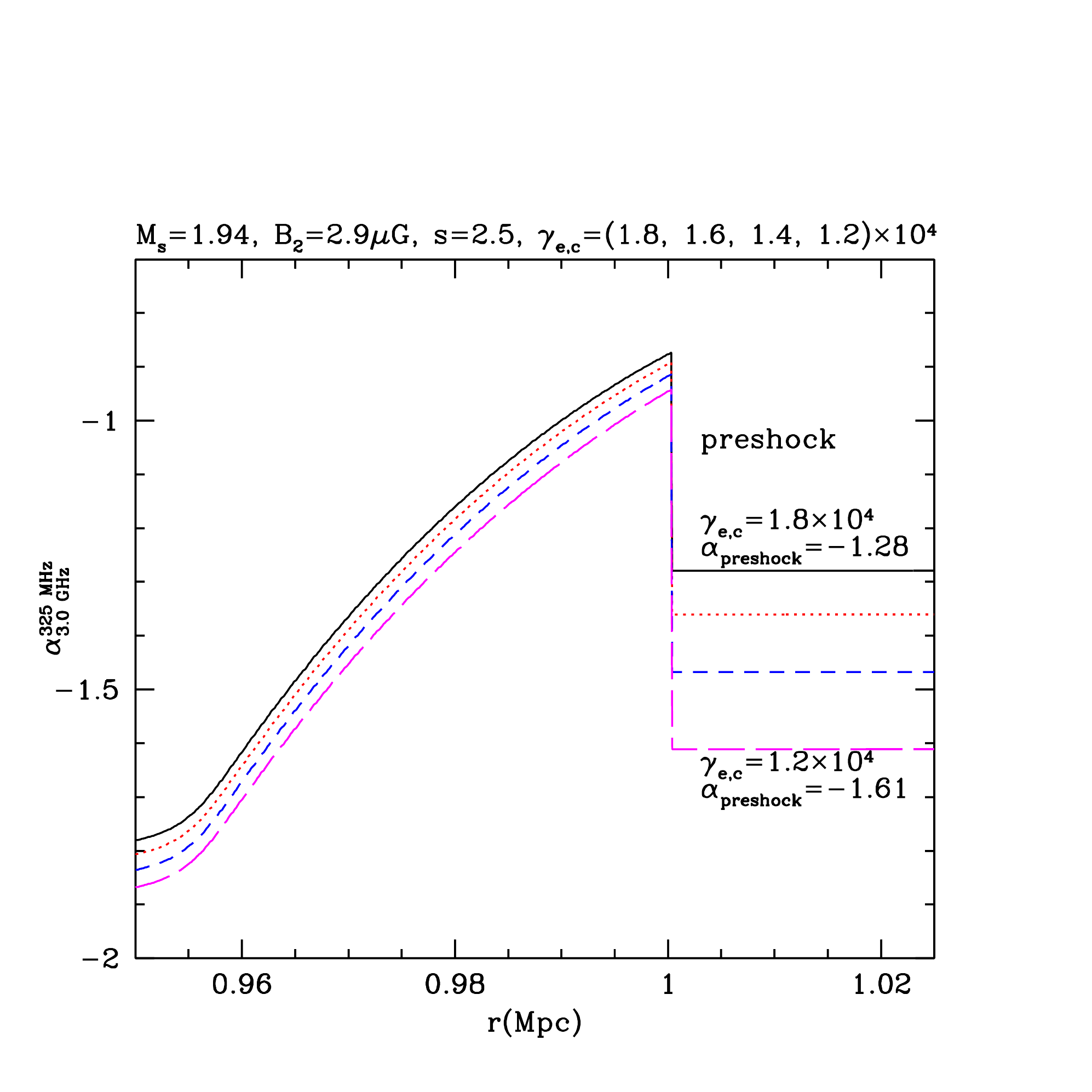}     
            \end{center}
       \caption{Simulated radio spectral index profiles across the relic radio for different values of  $\gamma_{e,c}$, to represent the continued steepening of the fossil plasma along the radio tail. In black, the profile at the relic's eastern end  with $\gamma_{e,c} = 1.8 \times 10^4$. The magenta (long dashed) line represents the radio spectral index on the relic's western end ($\gamma_{e,c} = 1.2 \times 10^4$). Intermediate values of $\gamma_{e,c}$ are shown in red (dotted; $\gamma_{e,c} = 1.6 \times 10^4$) and  blue (dashed; $\gamma_{e,c} = 1.4 \times 10^4$). It can be seen that the re-acceleration process at the shock partly ``erases'' the spectra index gradient ($\Delta\alpha=0.3$) and reduces it to ($\Delta\alpha=0.1$) after re-acceleration, creating a relatively uniform spectral index along the relic's length.}
       \label{fig:profilesteeperfossiltrend}
 \end{figure*}
\clearpage

\subsection{Subaru and Keck observations.}
\label{sec:SubaruKeck2}

\begin{figure}[h!]
\begin{center}
\includegraphics[width=0.8\columnwidth]{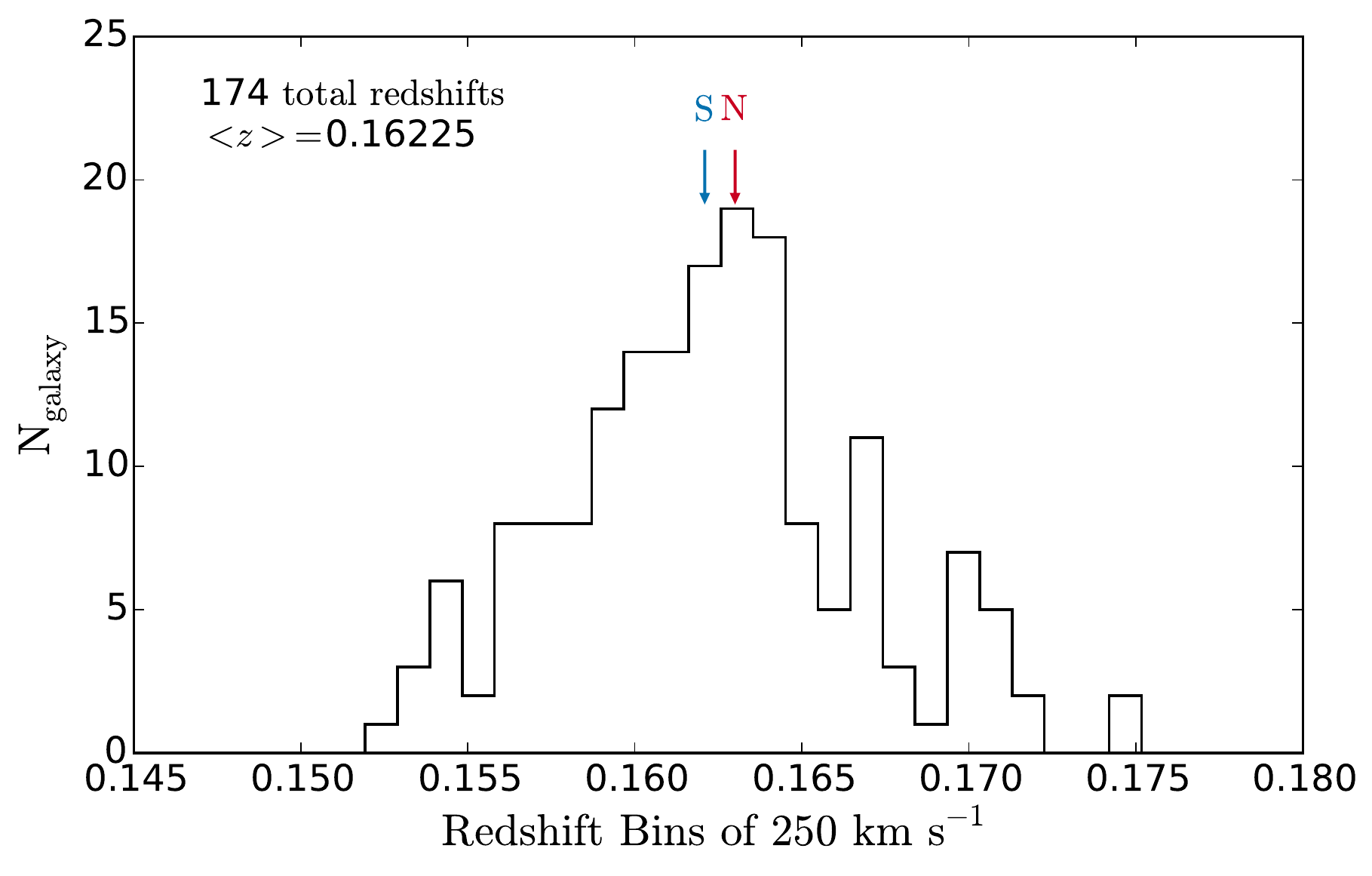}
\caption{Redshift distribution of the Keck DEIMOS high quality (Q $\geq$ 3) galaxy spectroscopic redshifts within the cluster redshift range. The north (A3411) brightest-cluster-galaxy (BCG) redshift is indicated by the red arrow, and the southern (A3412) BCG redshift is indicated by the blue arrow.
}
\label{fig:hist_deimos_spec}
\end{center}
\end{figure}

\subsubsection{Subcluster identification}
\label{sec:Subclusters}

To determine which galaxies are members of the Abell 3411 and 3412 subclusters, we utilize both spectroscopic and red sequence cluster member selection methods. 
The spectroscopic sample has the advantage of being a purer sample and the precise redshifts are a necessity for many of the following analyses.
While the red sequence sample is not as pure, it is more complete and it is not subject to the under sampling bias that affects the spectroscopic sample; thus it is advantageous for some analyses.

For the red sequence sample, we selected cluster members by first separating galaxies from stars using a size-magnitude relation based on the Subaru imaging (note that this separation is much more effective for the Subaru imaging than for the INT/WFC imaging described in the Methods section of the main article). We identified the red sequence in a color-magnitude diagram with $g-r$ as the color and $r$ as the magnitude. We used the spectroscopic survey to inform the red sequence selection (see Supplementary Fig.~\ref{fig:ColorMagDiag}) and estimate the purity of the red sequence selection. For the spectroscopic objects, we find that 196 lie within the red sequence selection bounds. Of these 78\% are cluster galaxies, 0 are foreground galaxies, 13\% are background galaxies, and 9\% are stars.

\begin{figure*}[h!]
\begin{center}
\includegraphics[width=\columnwidth]{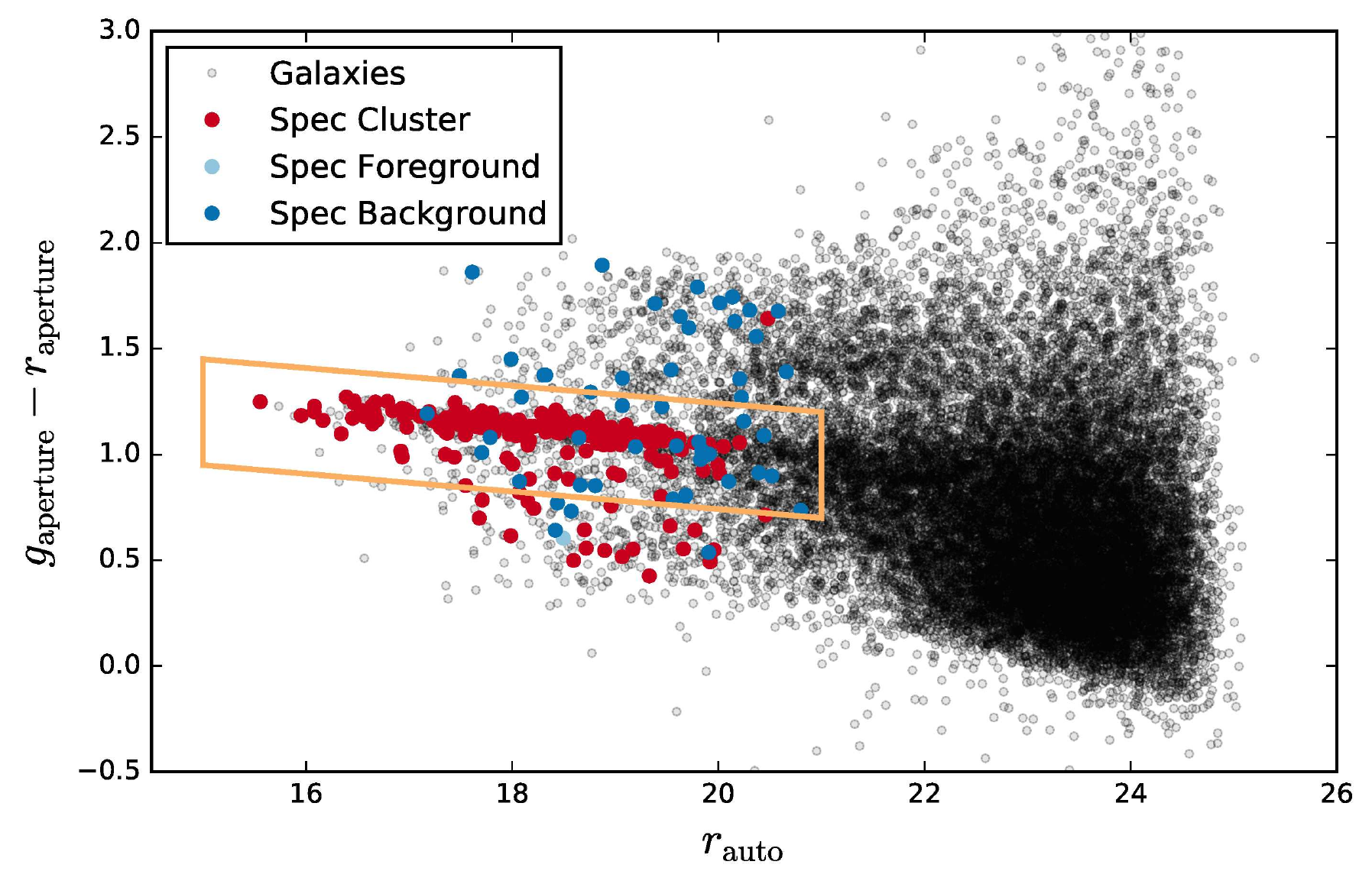}
\caption{
Color-magnitude diagram of galaxies within a 15$\arcmin$ radius of the system center, based on Subaru $g$ and $r$ magnitudes.
Spectroscopic cluster (red), foreground (light blue), and background (dark blue) galaxies are overlaid.
The red sequence selection region is outlined in orange.
}
\label{fig:ColorMagDiag}
\end{center}
\end{figure*}

No one subcluster identification method is optimal for all types of subcluster configurations\cite{1996ApJS..104....1P,2012A&A...540A.123E}, so we employ three separate complementary methods of subcluster identification. The first is a projected galaxy number/luminosity over-density analysis, the second is an Expectation-Maximization Gaussian Mixture Model (EM-GMM) clustering analysis, and the third is a Markov Chain Monte Carlo Gaussian Mixture Model (MCMC-GMM) clustering analysis.

We generated red sequence galaxy number density and luminosity density maps and bootstrapped to estimate the projected separation between the peaks of the two subclusters. From Fig.~1 (main article), it is clear that there are two dominant subclusters in projected space (corresponding to A3411 and A3412), but there are also small substructures to the north of A3411 and one northeast of A3412. We find that the corresponding spectroscopic cluster galaxy density maps agree well (see Supplementary Fig.~\ref{fig:speczdensity}), except that they do not show the smaller substructures since these lie outside the DEIMOS spectroscopic survey area.

\begin{figure}[h!]
\begin{center}
\includegraphics[width=1.0\columnwidth]{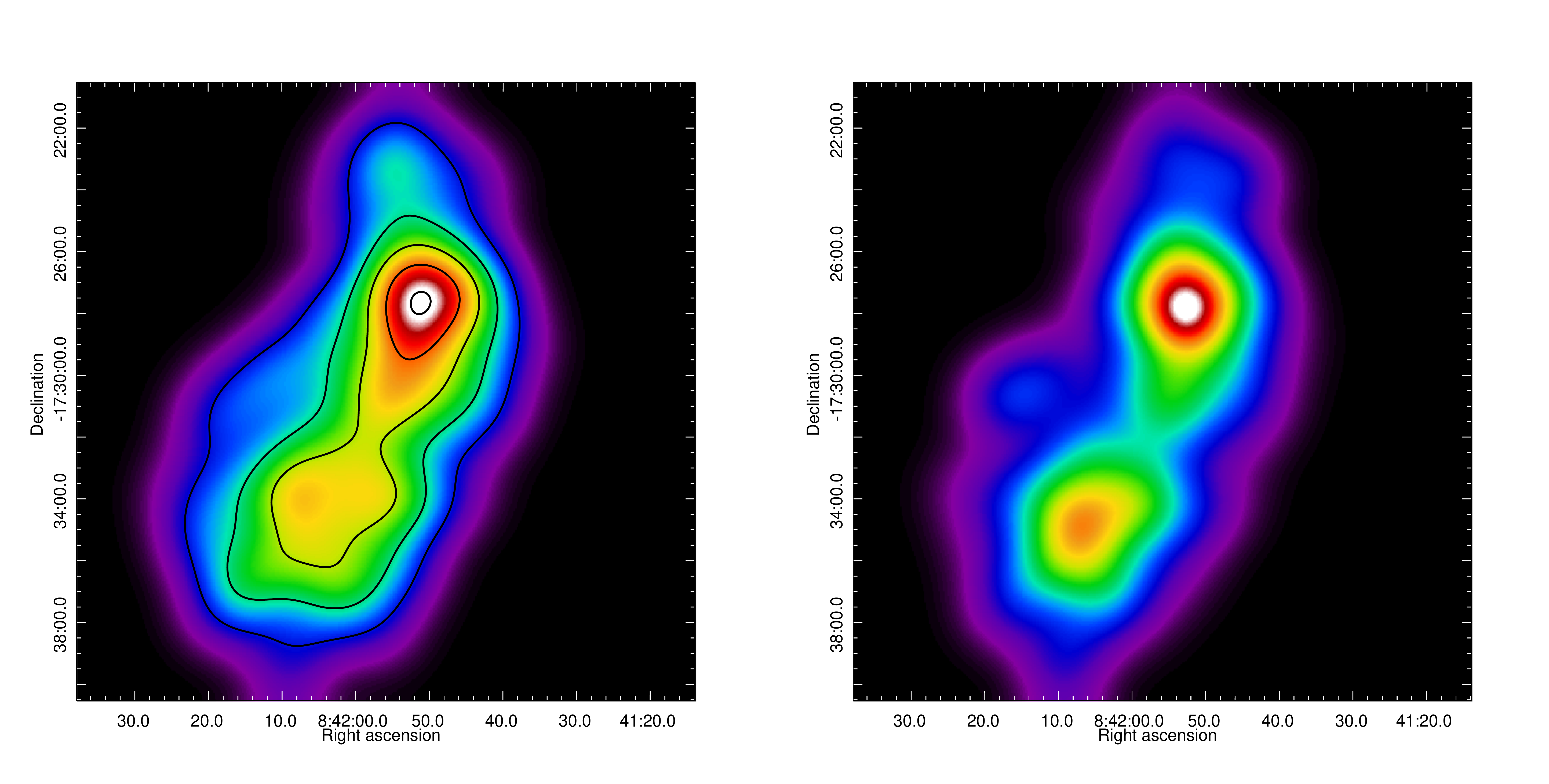}
\caption{
The spectroscopic galaxy density maps discussed in \S\ref{sec:Subclusters}. {Left:} The cluster spectroscopic sample number density map. The cluster spectroscopic number density contours (black) begin at 20\,galaxies\,Mpc$^{-2}$ and increase linearly with increments of 20\,galaxies\,Mpc$^{-2}$. {Right: } The cluster spectroscopic luminosity density map. Each map has a linear scaling with black being fewer and white being more galaxies\,Mpc$^{-2}$. Each map has been convolved with a 60\arcsec~Gaussian smoothing kernel.}
\label{fig:speczdensity}
\end{center}
\end{figure}

For the EM-GMM, we use \emph{scikit-learn}'s GMM package and apply it to the three-dimensional galaxy distribution (right ascension, declination, and redshift) of the spectroscopic sample of cluster members.
We consider mixtures of 1 to 7 multivariate Gaussian components  with a diagonal (\emph{diag}) covariance structure, where each Gaussian component has an uncorrelated covariance structure.
While this covariance structure prevents angled elliptical halos, we find that the other covariance structures in \emph{scikit-learn}'s GMM package offer too much flexibility (e.g., the \emph{full}, where each Gaussian component can have a different unstructured covariance, results in unphysical solutions where some subclusters have substantial coherent rotations).
For each number of components ($n$), we calculate the BIC (Bayesian information criterion) and use this to infer the optimal number of subclusters.
We plot these results as, 
\begin{equation}
\Delta\mathrm{BIC}_{n} = \mathrm{BIC}_{n} -  \min\left(\mathrm{BIC}_{n}|n\in\mathbb{Z}_{1...7}\right)
\end{equation}
where $\mathbb{Z}_{1...7}$ is the set of integers from 1 to 7.
For convenience of interpretation, we color-code regions of the $\Delta$BIC plot according to the broad model comparison categories suggested by\cite{Kass:1995}.

We find that, of the models considered, the two-component Gaussian model is the most favored one (see Supplementary Fig.~\ref{fig:3D-gmmselect}).
In Supplementary Fig.~\ref{fig:3D-gmmtriangle}, we plot the three-dimensional distribution of the spectroscopic cluster members and their most likely subcluster membership assignment for this lowest-BIC model.
For the projected one-dimensional distributions, we plot the marginalized Gaussian components for the lowest-BIC model (dashed lines).
For the projected two-dimensional distributions we plot marginalized 68\% confidence ellipses of the lowest-BIC model Gaussian components.
The EM-GMM identifies two subclusters that correspond to A3411 and A3412.
We also employed a more sophisticated MCMC-GMM analysis which is similar to the EM-GMM, but facilitates the implementation of informative priors on the model parameters (e.g., location of the halos, typical cluster scales, and typical cluster ellipticities).
For this system, the results of the two GMM analyses produce nearly identical results.

\begin{figure}[h!]
\begin{center}
\includegraphics[width=0.6\columnwidth]{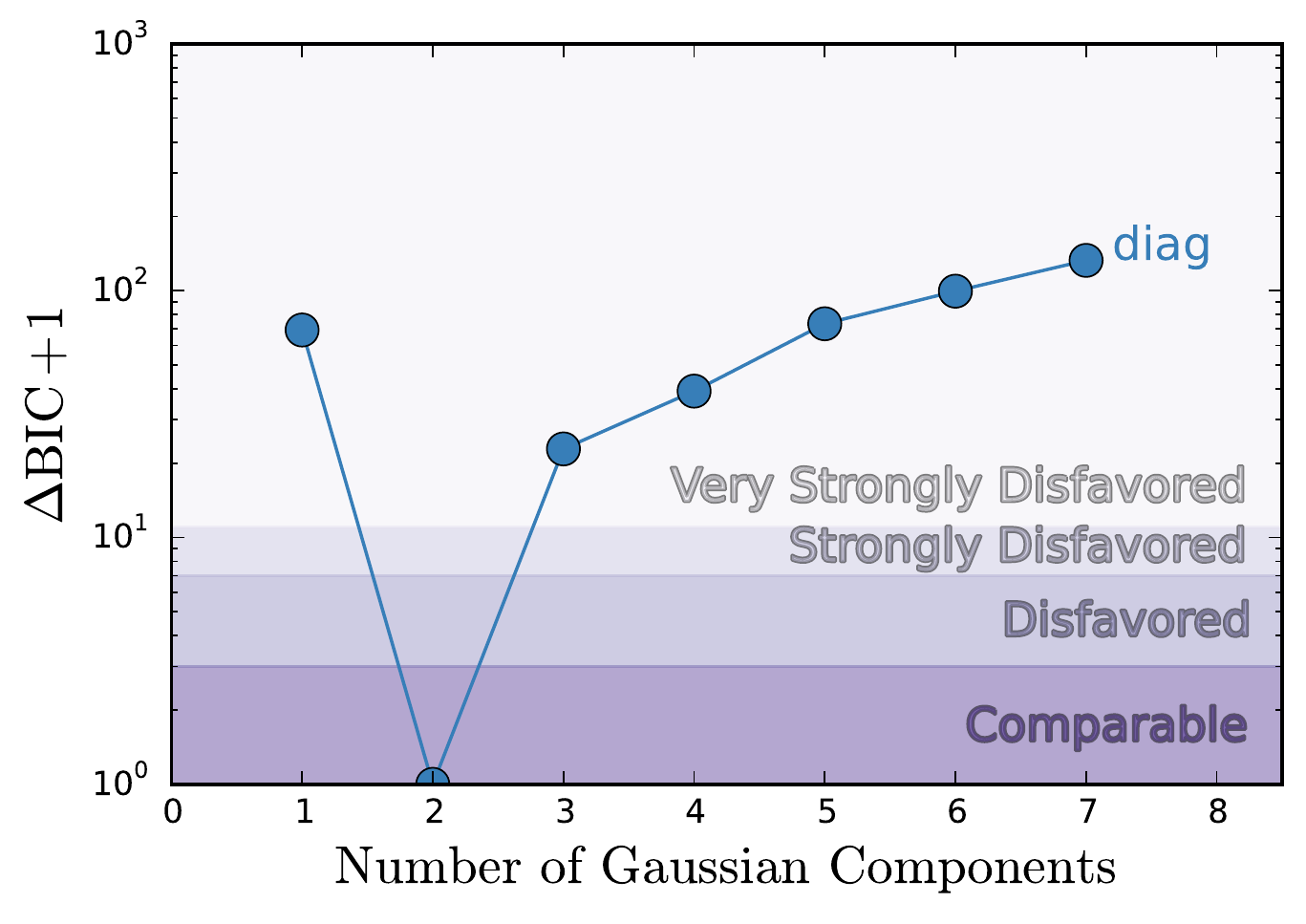}
\caption{
A $\Delta$BIC plot comparing GMM fits to the three-dimensional (right ascension, declination, and redshift) distribution of all the cluster member spectroscopic galaxies, with varying number of Gaussian components and a diagonal covariance type. 
The purple shaded regions roughly denote how a given model compares with the model that has the lowest BIC score.
The most favored is a 2 component model with \emph{diag} covariance structure.
The two components closely correspond to A3411 and A3412.
}
\label{fig:3D-gmmselect}
\end{center}
\end{figure}

\begin{figure}[h!]
\begin{center}
\includegraphics[width=\columnwidth]{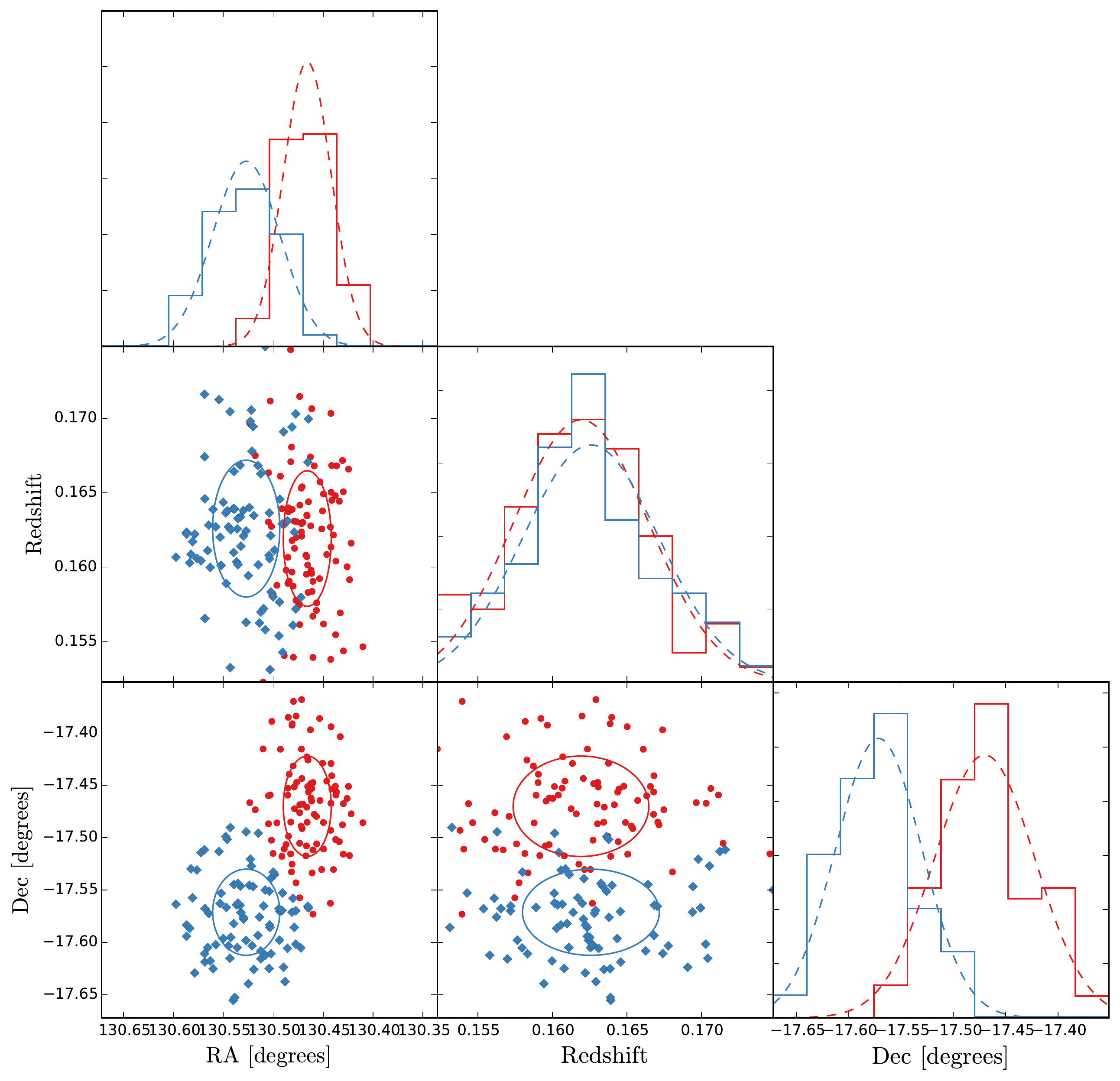}
\caption{
Three-dimensional distribution of the spectroscopic cluster members (\S\ref{sec:Subclusters}) and their most likely subcluster membership assignment for the lowest-BIC EM-GMM (see Supplementary Fig.~\ref{fig:3D-gmmselect}).
For the projected one-dimensional distributions, we plot the marginalized Gaussian components for the lowest-BIC model (dashed lines). For the projected two-dimensional distributions we plot projected ellipses that encompass  $\sim68\%$ of the corresponding members in the lowest-BIC  model Gaussian components.
}
\label{fig:3D-gmmtriangle}
\end{center}
\end{figure}

\subsubsection{Dynamical analysis and masses.}
\label{sec:dynamics}

To investigate the redshift and velocity dispersion of each subcluster, we subdivide the cluster spectroscopic sample (\S\ref{sec:SubaruKeck2}) according to MCMC-GMM analysis (\S\ref{sec:Subclusters}) maximum likelihood membership probabilities.
The redshift distributions of each of these selections are shown in Supplementary Fig.~\ref{fig:SubclustSpecProps}.

We estimate each subcluster's redshift and velocity dispersion using the biweight-statistic and bias-corrected 68\% confidence limit\cite{1990AJ....100...32B} applied to 100,000 bootstrap samples of each subcluster's spectroscopic redshifts.
We find very similar redshifts for the A3411 and A3412 subclusters, $0.16194^{+0.00045}_{-0.00046}$ and $0.16226^{+0.00047}_{-0.00048}$, respectively.
These translate to a relative line-of-sight (LOS) velocity difference in the frame of the cluster of  $v_\mathrm{\rm{A3411}}-v_\mathrm{\rm{A3412}}=-80\pm 170\,\mathrm{km}\,\mathrm{s}^{-1}$. 
This suggests that either they are both nearly in the plane of the sky, have slowed as they near the apocenter of the merger, or a combination of the two. Comparing the relative redshift of each subcluster's BCG with respect to the median subcluster redshift, we find relative line-of-sight velocity differences of $v_\mathrm{\rm{A3411}}-v_\mathrm{\rm{A3411\,BCG}} = -270\pm120$\,km\,s$^{-1}$ and $v_\mathrm{\rm{A3412}}-v_\mathrm{\rm{A3412\,BCG}} = +40\pm120$\,km\,s$^{-1}$.

We find similar velocity dispersions for the northern (A3411) and southern (A3412) subclusters, $1110^{+100}_{-80}\,\mathrm{km}\,\mathrm{s}^{-1}$ and $1190^{+100}_{-90}\,\mathrm{km}\,\mathrm{s}^{-1}$, respectively.
Converting these velocity dispersions into $M_{200}$ mass estimates using the\cite{2008ApJ...672..122E} scaling relation, we estimate masses of $14^{+4}_{-3}\times 10^{14}$\,M$_\odot$ and  $18^{+5}_{-4}\times 10^{14}$\,M$_\odot$ for the A3411 and A3412 subclusters, respectively.
Velocity dispersion mass estimates can be biased high in disturbed systems, however this bias primarily exists when the merging halos are separated by $\lesssim R_{200}$ in the $\sim1$\,Gyr pre/post collision, and is approximately a factor of three worse for mergers occurring along the line-of-sight\cite{1996ApJS..104....1P}.
We expect that the velocity dispersion mass estimates of A3411 and A3412 are not significantly biased because they have a projected separation of $1.4\pm0.2$\,Mpc,  the dynamics analysis suggests (see the next paragraph) that the time-since-collision (TSC) is $>0.9$\,Gyr, and the angle of the merger axis with respect to the plane-of-the-sky ($\alpha$) is $23\degr^{+18}_{-14}$. Note that $\alpha$ is largely insensitive to the mass of the subclusters\cite{2012ApJ...747L..42D}.

For the subcluster two-body dynamical analyses, we use the MCMAC Monte Carlo merging cluster dynamics analysis method \cite{2013ApJ...772..131D}.
The MCMAC program takes PDF's (probability density functions) for the mass, redshift, and projected separation of two subclusters as input.
We use the aforementioned redshifts, velocity dispersion based mass estimates, and red sequence luminosity density based projected separation estimate.
Additionally, we incorporate a 20\% radio polarization fraction of the relic as a prior that $\alpha<48\degr$\cite{1998A&A...332..395E,2015MNRAS.453.1531N}.
We summarize the geometric and dynamic parameter constraints in Supplementary Table~\ref{tbl:dynresultparam}.
While there is some ambiguity in whether the two subclusters are on an outbound trajectory or if they have reached their apocenter and are on a return trajectory, we find that the system is near apocenter and the time-since-collision for the outbound case ($TSC_0$) is close to that of the return case ($TSC_1$), see Supplementary Table~\ref{tbl:dynresultparam}.
Supplementary Fig.~\ref{fig:SubclustSpecProps} (right panel) provides an example of the marginal posterior parameter constraints from the dynamics analysis and the impact of the polarization prior.

We thus conclude that core passage for the A3411-A3412 merger event occurred about $\sim 1$~Gyr ago and the merger event is seen relatively close (9\degr -- 41\degr) to edge on.

\begin{figure}[h!]
\begin{center}
\includegraphics[width=0.49\columnwidth]{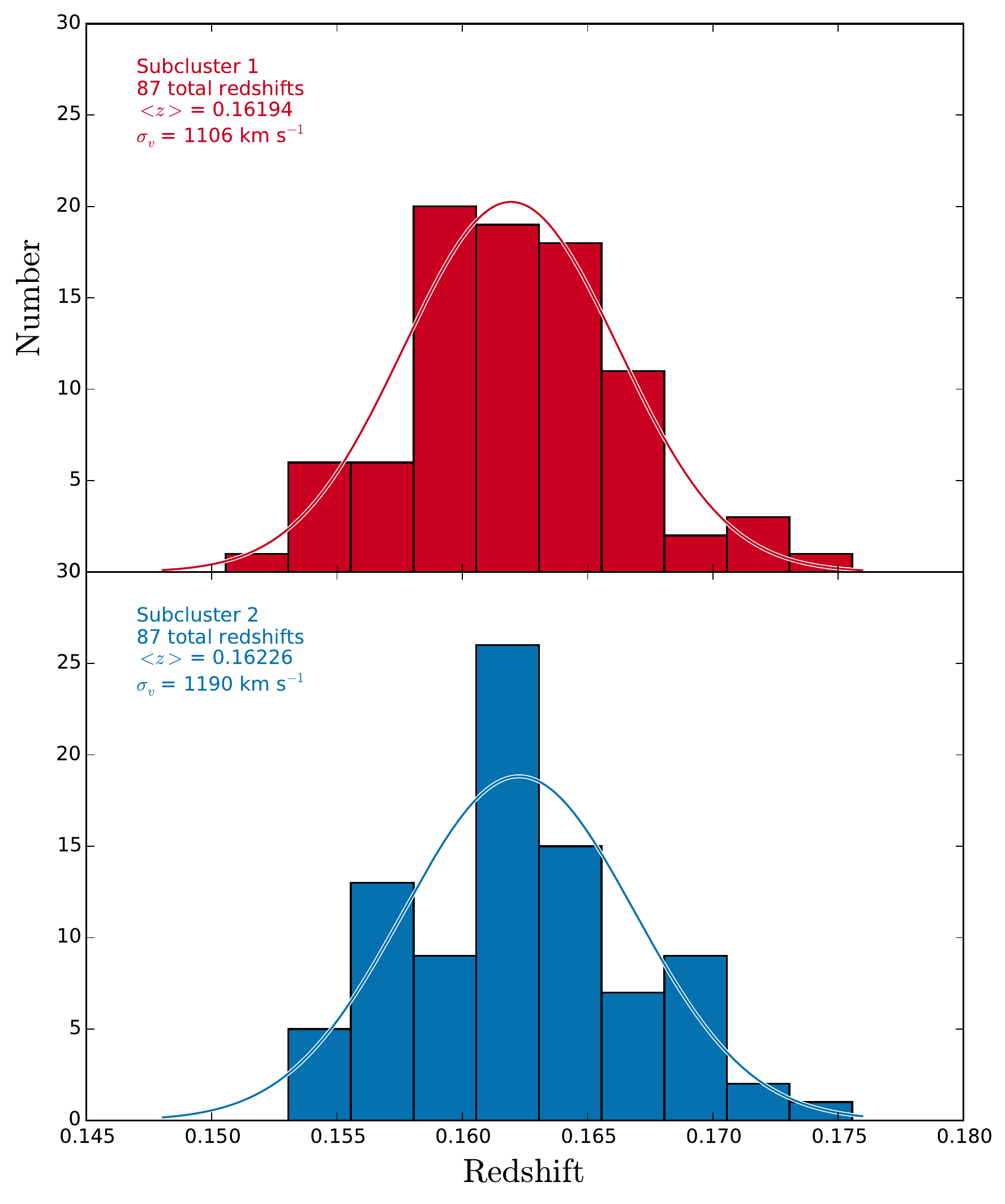}
\includegraphics[width=0.49\columnwidth]{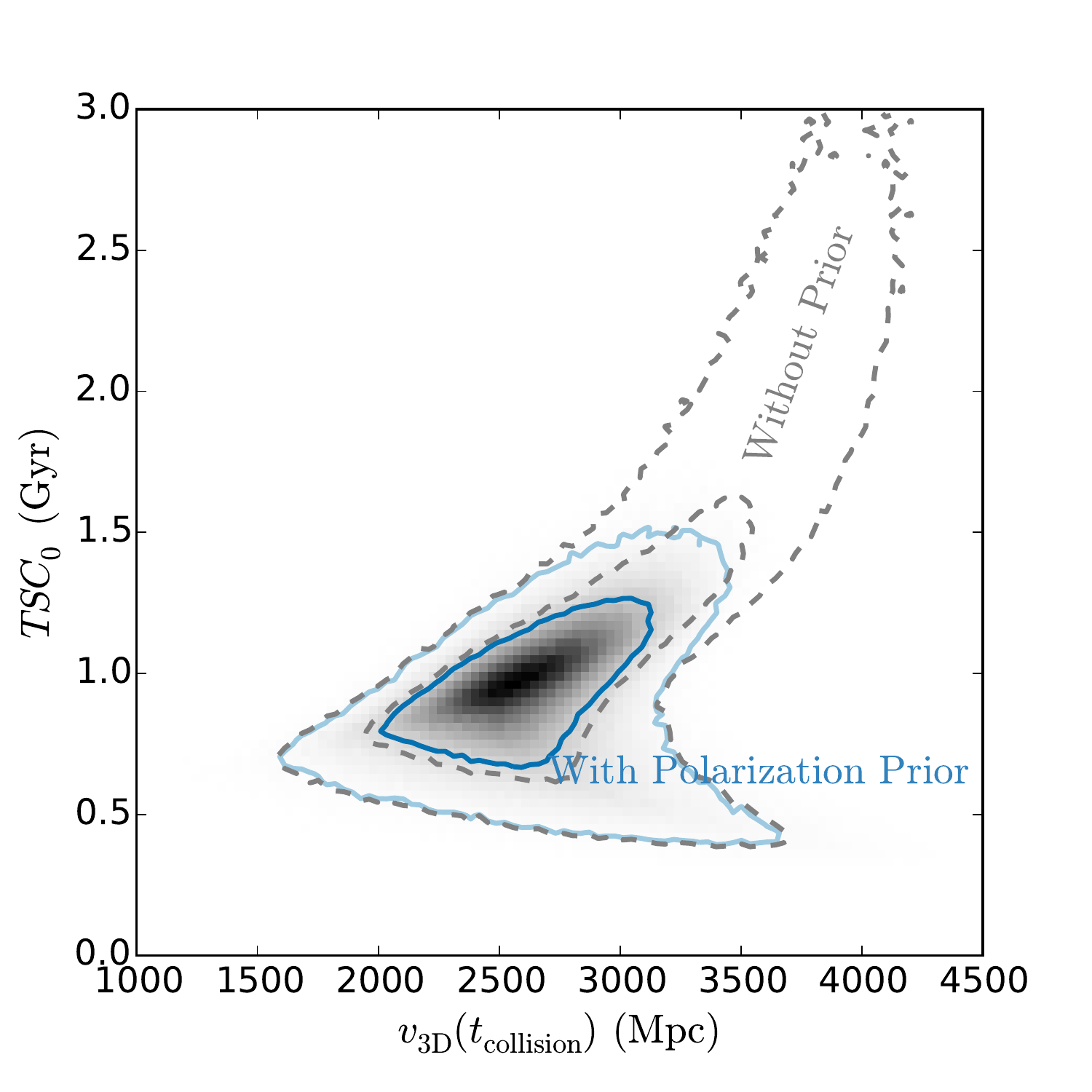}
\caption{
Left: Redshift distributions for the two subclusters, following the label and color scheme of Supplementary Fig.~\ref{fig:3D-gmmtriangle}.
Redshift locations and velocity dispersions are listed in the upper left of each subpanel. 
The galaxy membership of each subcluster is based on the MCMC-GMM analysis (see \S\ref{sec:Subclusters}).
Right: Marginal posterior of the relative three-dimensional subcluster velocity at the time of collision $v_\mathrm{3D} (t_\mathrm{col})$ and the time since collision given their observed state, inferred from the Dawson (2013) dynamics analysis. Dark and light blue contours represent the 68 and 95\% confidence regions, respectively, for the case with the radio relic polarization prior based on a polarization fraction of 20\%. The gray dashed contours represent the 68 and 95\% confidence regions, respectively, for the case of no prior.
}
\label{fig:SubclustSpecProps}
\end{center}
\end{figure}

\begin{table}[h!]
\caption{A3411-3412 dynamic analysis parameter estimates including the radio relic polarization prior\label{tbl:dynresultparam}}
\begin{center}
\begin{tabular}{lcccc}
\hline
\hline
Parameter\tablenotemark{a}     & Units  & Location\tablenotemark{b}  & 68\% LCL--UCL\tablenotemark{c} & 95\% LCL--UCL\tablenotemark{b}\\
\hline
M$_{200_1}$	&	$10^{14}$\,M$_\odot$	&	14	&	11	--	19	&	7	--	22	\\
M$_{200_2}$	&	$10^{14}$\,M$_\odot$	&	18	&	13	--	22	&	9	--	27	\\
$z_1$	&		&	0.1620	&	0.1615	--	0.1624	&	0.1612	--	0.1628	\\
$z_2$	&		&	0.1622&	0.1619	--	0.1627	&	0.1614	--	0.1632	\\
$d_{\rm proj}$		&	Mpc	&	1.4	&	1.2	--	1.6	&	0.9	--	1.9	\\
$\alpha$	&	degree	&	23	&	9	--	41	&	2	--	47	\\
$d_{\rm 3D}$	&	Mpc	&	1.5	&	1.2	--	1.9	&	0.9	--	2.3	\\
$d_{\rm max}$	&	Mpc	&	1.6	&	1.3	--	1.9	&	0.9	--	2.5	\\
$v_{\rm 3D}(t_{\rm obs})$	&	km\,s$^{-1}$	&	330	&	50	--	760	&	50	--	1880	\\
$v_{\rm 3D}(t_{\rm col})$ &	km\,s$^{-1}$	&	2600	&	2400	--	3000	&	2000	--	3400	\\
$TSC_0$	&	Gyr	&	0.9	&	0.9	--	1.3	&	0.7	--	1.7	\\
$TSC_1$	&	Gyr	&	1.2	&	1.0	--	1.6	&	0.9	--	2.5	\\
$T$	&	Gyr	&	2.2	&	1.9	--	2.6	&	1.6	--	3.3	\\
\hline
\end{tabular}
\end{center}
$^{a}${$M_{200}$ mass; $z$ redshift; $d_{\rm{proj}}$ projected subcluster separation; $\alpha$  merger axis' angle with respect to the plane-of-the-sky; $d_{\rm 3D}$ 3-D subcluster separation; $d_{\rm max}$ 3-D maximum subcluster separation after core passage;  $v_{\rm 3D}(t_{\rm obs})$ current 3-D velocity difference between the subclusters; $v_{\rm 3D}(t_{\rm col})$ subcluster velocity difference at the time of core passage; $TSC_0$ time-since-collision (time after core passage) for the outbound case; $TSC_0$ time-since-collision for the return case; $T$ time between collisions. See\cite{2013ApJ...772..131D} for more details on the these quantities.}\\
$^{b}${Biweight-statistic location\cite{1990AJ....100...32B}.} \\
$^{c}${Bias-corrected lower and upper confidence limits, LCL and UCL respectively\cite{1990AJ....100...32B}.} 
\end{table}

\clearpage

\begin{addendum}
 \item[Correspondence] Correspondence and requests for materials should be addressed to R.J.W. \\(rvanweeren@cfa.harvard.edu).
 
 \item [Acknowledgements] We thank the anonymous referees for comments. Support for this work was provided by the National Aeronautics and Space Administration through Chandra Award Number GO5-16133X issued by the Chandra X-ray Observatory Center, which is operated by the Smithsonian Astrophysical Observatory for and on behalf of the National Aeronautics Space Administration under contract NAS8-03060. We thank the staff of the GMRT who have made these observations possible. 
The GMRT is run by the National Centre for Radio Astrophysics of the Tata Institute of Fundamental Research. 
The National Radio Astronomy Observatory is a facility of the National Science 
Foundation operated under cooperative agreement by Associated Universities, Inc. Based on observations obtained at the Southern Astrophysical Research (SOAR) telescope, which is a joint project of the Minist\'erio da Ci\^encia, Tecnologia, e Inova\c{c}\~ao (MCTI) da Rep\'ublica Federativa do Brasil, the U.S. National Optical Astronomy Observatory (NOAO), the University 
of North Carolina at Chapel Hill (UNC), and Michigan State University (MSU). Based on data collected at Subaru Telescope, which is operated by the National Astronomical Observatory of Japan. Part of this work performed under the auspices of the U.S. DOE by LLNL under Contract DE-AC52-07NA27344. Some of the data presented herein were obtained at the W.M. Keck Observatory, which is operated as a scientific partnership among the California Institute of Technology, the University of California and the National Aeronautics and Space Administration. The Observatory was made possible by the generous financial support of the W.M. Keck Foundation. The Isaac Newton Telescope is operated on the island of La Palma by the Isaac Newton Group in the Spanish Observatorio del Roque de los Muchachos of the Instituto de Astrof\'isica de Canarias.
R.J.W. is supported by a Clay Fellowship awarded by the Harvard-Smithsonian Center for Astrophysics. V.M.P. acknowledges support for this work from grant PHY 14-30152; Physics Frontier Center/JINA Center for the Evolution of the Elements (JINA-CEE), awarded by the US National Science Foundation. D.R. was supported by the National Research Foundation of Korea through grant 2016R1A5A1013277. H.K. was supported by the National Research Foundation of Korea through grant 2014R1A1A2057940. R.M.S. acknowledges CAPES (PROEX), CNPq, PRPG/USP, FAPESP and INCT-A funding. M.J.J. acknowledges the support from NRF of Korea to CGER. D.S. acknowledges financial support from the Netherlands Organisation for Scientific research (NWO) through a Veni fellowship. G.A.O. is supported by NASA through Hubble Fellowship grant \#HST-HF2-51345.001-A, awarded by the Space Telescope Science Institute, which is operated by the Association of Universities for Research in Astronomy, Inc., for NASA, under contract NAS5-26555.

\item [Author Contributions] R.J.W. coordinated the research, wrote the manuscript, reduced the VLA data, and led the Chandra observing proposal. F.A.-S., K.F., and G.A.O. performed the Chandra data reduction and worked on the X-ray surface brightness profile fitting. H.K. and D.R. carried out the re-acceleration modeling. M.B., W.R.F., and C.J. helped with the interpretation of the radio and X-ray results and provided extensive feedback on the manuscript. C.J. led the GMRT observing proposal. D.V.L. obtained the GMRT observations and carried out the GMRT data reduction.  V.M.P. and R.M.A. obtained the SOAR observations and performed the corresponding data reduction. D.S. and A.S. provided the INT observations and reduced the data. W.A.D. carried out the dynamical modeling of the merger event. W.A.D, N.G. and M.J.J. obtained the Keck and Subaru observations and reduced the data. D.W. helped with the interpretation of the dynamical modeling and led the Keck and Subaru observing proposals. R.P.K. assisted with the writing of the Chandra observing proposal.

\end{addendum}

\end{document}